\documentclass[twocolumn]{aastex701}
\usepackage{gensymb}
\usepackage{bm}
\usepackage{amsmath}
\newcommand{\pcc}{{\rm cm}^{-3}}

\shorttitle{Gravity-driven longitudinal flows in filaments on angular momentum transport to cores}
\shortauthors{Arroyo-Chávez et al.}

\begin{document}

\title{Effect of gravity-driven longitudinal flows in filaments on angular momentum transport to embedded cores}

\author[orcid=0000-0002-7082-0587]{Griselda Arroyo-Chávez}
\altaffiliation{Steward Observatory}
\affiliation{Steward Observatory, University of Arizona 933 North Cherry Avenue, Tucson AZ, 85721, USA}
\email[show]{arroyochavezg@arizona.edu}  

\author[orcid=0000-0000-0000-0002]{Shuo Kong} 
\altaffiliation{Steward Observatory}
\affiliation{Steward Observatory, University of Arizona 933 North Cherry Avenue, Tucson AZ, 85721, USA}
\email{shuokong@arizona.edu}

\author[orcid=0000-0002-1424-3543]{Enrique V\'azquez-Semadeni}
\altaffiliation{IRyA UNAM}
\affiliation{Instituto de Radioastronom\'ia y Astrof\'isica, PO Box 3-72. 58090, Morelia, Michoac\'an, M\'exico}
\email[]{e.vazquez@irya.unam.mx} 

\begin{abstract}
Different models of filament formation predict distinct patterns of angular momentum redistribution toward embedded cores, set by the underlying velocity-field structure, which can set the initial conditions for a preferential orientation between protostellar outflows and filaments. However, the absence of a dominant alignment in observations keeps this connection open to debate. We investigate whether gravity-driven longitudinal flows along filaments can redistribute angular momentum (AM) toward collapse centers. To this end, we analyze the distributions of 3D and 2D-projected angles between sink angular momentum vectors and host filament orientations in an SPH simulation of giant molecular cloud and filament formation. We also characterize the filament velocity field by measuring the angles between SPH particle velocity vectors and filament axes, and the degree of convergent flow toward filament density peaks. No preferred alignment between the sinks' AM and the filament direction is found at early evolutionary stages, neither in 3D nor in 2D. Later, however, a predominantly perpendicular configuration emerges in 3D. Tracking individual sinks indicates that this alignment is not primordial but develops as gravity strengthens. In individual filaments, the onset of perpendicular alignment coincides with the development of convergent longitudinal flows. Finally, we estimate the minimum excess fraction of intrinsically perpendicular 3D orientations needed for a projected sample to retain a detectable perpendicular alignment in 2D as a function of sample size, and the minimum sample size for the trend to be detectable in projection, which is at least $\sim 35$ measurements, making conclusions from limited 2D samples potentially misleading.

\end{abstract}


\section{Introduction} 
\label{sec:Introduction}

In recent years, star formation has increasingly been studied not as an isolated event, but as a consequence of the complex interplay between phenomena operating across multiple scales \citep{BP+12,Colombo+2015,VS+19}. In particular, clumps and dense cores have been shown to reside within a hierarchical network of filaments that channels the gas toward local collapse centers \citep{Schneider.Elmegreen1979,Bally+1987,Andre+2010,Molinari+2010,Hacar+2013,Andre+2014,Planck+2016,Hacar+2016,Hacar+2018}. The dynamical study of the gas around these filaments has revealed the presence of multiple velocity components, including perpendicular accretion onto them  \citep{Palmeirim+2013,Chen+2020}, longitudinal flows along their main axis \citep{Myers+2009,Kirk+2013,Peretto+2014,Gomez.Vazquez14,Ren+2021,Rawat+2024}, and rotation in a few examples \citep{Hsieh+2021,Alvares-Gutierrez+2021,Saliinas+2025}. These gas flows can directly impact the star formation process within filaments, particularly by regulating accretion onto dense cores and subsequently from the cores onto protostars \citep{Anderson+2021,Yang+2023}. Additionally, several studies have quantified both the direction and magnitude of the angular momentum of cores, considering the environment in which they form, including filamentary structures. These measurements have often been interpreted in terms of angular momentum acquisition mechanisms, including turbulence and accretion \citep[e.g.][]{Misugi+2019,Misugi+2023,Misugi+2023_2, Kuznetsova+2019}. In turn, as a consequence of this accretion, feedback in the form of jets and outflows play a decisive role in local dynamics and the transport of momentum and energy to the surrounding environment, affecting the internal dynamics of their host filaments \citep{Arce+2007}. 

The orientation of protostellar outflows with respect to their host filaments has motivated several observational studies of the connection between filament-scale gas dynamics and the angular momentum of star-forming material. The observed distributions of projected angles between outflows and filaments tend to be mostly random \citep{Stephens+2017,Baug+2020}, with some particular cases of perpendicular alignments \citep{Kong+2019,Feddersen+2020} and, less frequently, parallel ones \citep{Green+2024}.

\begin{figure}
\centering \offinterlineskip
\includegraphics[width=0.9\linewidth]{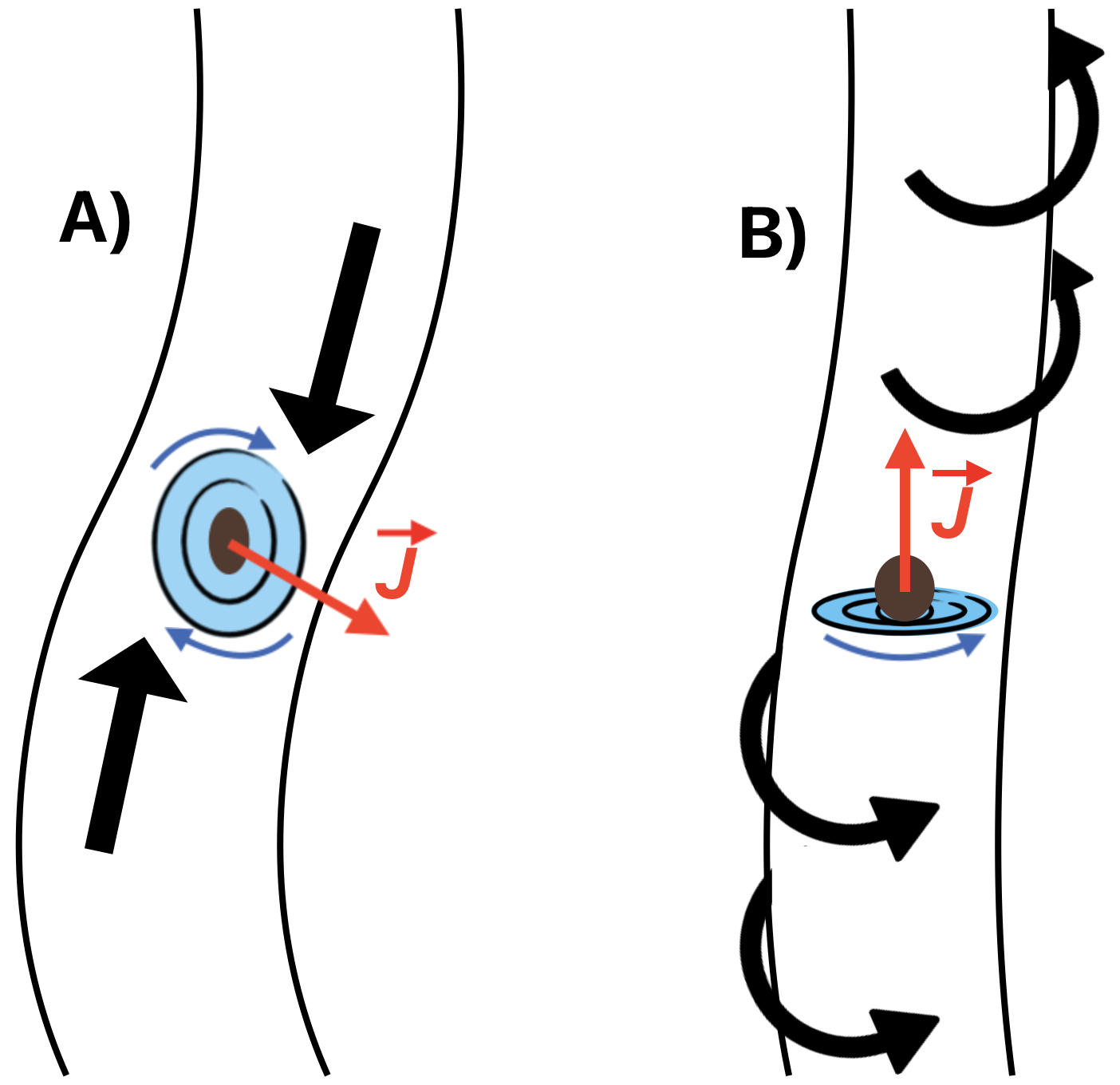}
 \caption{{\it A)} Prediction of perpendicular OFA in the model presented by \citet{Anathprindika.Withworth2008}. The misalignment of gas inflows toward the hub generates a net torque that induces rotation, such that the resulting angular momentum vector ($\vec{J}$) of the hub—and subsequently the outflows direction—is oriented perpendicular to the major axis of the filament. {\it B)} Prediction of pparallel OFA in the model presented by \citet{Banerjee.Pudritz.Anderson2006}. The core inherits the filament’s rotation about its major axis, such that $\vec{J}$ is parallel to the filament axis.}
 \label{fig:models}
\end{figure}

Two theoretical models of core formation offer insight into how filament-scale gas flows may set the angular momentum direction of embedded fragments, under the idealized assumption that this angular momentum may later influence the orientation of the protostellar system. For instance, \citet{Anathprindika.Withworth2008} propose that filaments can form through the cooling of post-shock gas produced in cloud-cloud collisions. In this scenario, the orbital angular momentum of the colliding clouds is oriented perpendicular to the filaments that form within the compressed layer and evolve dynamically as the collision progresses. As the layer tumbles, the filaments move and deform accordingly, leading to increasing misalignment between the longitudinal gas flows that converge along them. This asymmetry generates a net torque on the central hub, imparting rotation. Consequently, the angular momentum vector ($\vec{J}$, see panel A in Fig. \ref{fig:models}) of the core becomes oriented perpendicular to the filament, producing a perpendicular outflow-filament alignment (OFA)\footnote{It is important to clarify that, from this point onward, our reference to this scenario concerns specifically the effect of misaligned longitudinal flows converging toward a collapse center in filaments. We do not, however, make any assumptions regarding the primary origin of the filaments themselves, whether they arise from cloud-cloud collisions, large-scale converging flows, or turbulent compression in the diffuse medium.}. In contrast, \citet{Banerjee.Pudritz.Anderson2006} suggest that filaments arise from shear in turbulent oblique shocks, which impart an initial rotation aligned with the filament’s major axis. This rotation is inherited by the cores during fragmentation, naturally leading to a parallel OFA (see panel B in Fig. \ref{fig:models}).

The numerical simulation by \cite{Gomez.Vazquez14} of a head-on collision of two streams of warm atomic gas, appears more consistent with \citet{Anathprindika.Withworth2008}. In their simulation, filaments form in the compressed cold atomic layer by a combination of thermal and gravitational instabilities, and the filaments in turn develop local collapsing sites (cores and hubs) at their densest parts by gravitational instability. The gravitational attraction of these hubs and cores then generates a longitudinal flow along the filaments. Although these authors do not explicitly examine the angular momentum of the cores and hubs formed, their simulations reveal a delay between their formation and the onset of longitudinal flows toward them. Once established, these flows impart rotation, implying that those formed later may have a preferential angular momentum orientation compared to those formed earlier. This picture is further supported by the results of \citet{Tsukamoto.Machida2013}, who follow the evolution of angular momentum orientations in protostellar disks relative to their parent cores. They find that disk and core angular momenta are initially uncorrelated, but progressively evolve toward a preferentially parallel configuration, after sustained periods of accretion, despite the turbulent nature of the environment.

In this work, we use a numerical simulation of cloud and filament formation to investigate how gravity-driven longitudinal flows along filaments can torque the gas at convergence points and influence the angular momentum of unresolved collapsing fragments. Although the observational motivation comes partly from outflow-filament alignment studies, establishing a direct connection between the angular momentum measured here and the direction of protostellar outflows requires resolving the subsequent collapse to dense-core, disk, and launching scales, including magnetic fields. This lies beyond the scope of the present work. Furthermore, the mechanisms illustrated in Fig. \ref{fig:models} do not consider magnetic fields; accordingly, this study focuses exclusively on the gas dynamics within and around the filaments (see Sec. \ref{sec:Caveats} on caveats).

The structure of this paper is as follows. Section \ref{sec:Numerical data} describes the main characteristics of the simulation analyzed, followed by Section \ref{sec:Methods}, which details the methods and procedures employed in the analysis. Our results are presented in Section \ref{sec:Results}, with a discussion provided in Section \ref{sec:Discussion}. Finally, the main conclusions are summarized in Section \ref{sec:Conclusions}.

\section{Numerical data}
\label{sec:Numerical data}

We use the smoothed-particle hydrodynamics (SPH) code \textsc{Phantom} \citep{Price+2018}, to perform a purely hydrodynamic simulations of decaying turbulence including gravity (labeled HDG3). Further details regarding this simulation are presented by \citet{Arroyo-Chavez+2026}. The setup arises from combining the cluster \textsc{Phantom} setup preloaded in the public version derived from \cite{Bate_Bonnell_Bromm2003}, and the turbulent driving module based on the works of \citet{Price.Federrath2010} and \citet{Tricco.Price.Federrath2016,Tricco.Price.Laibe2017}.

\begin{figure*}
\centering
\centering \offinterlineskip
\includegraphics[width=\textwidth]{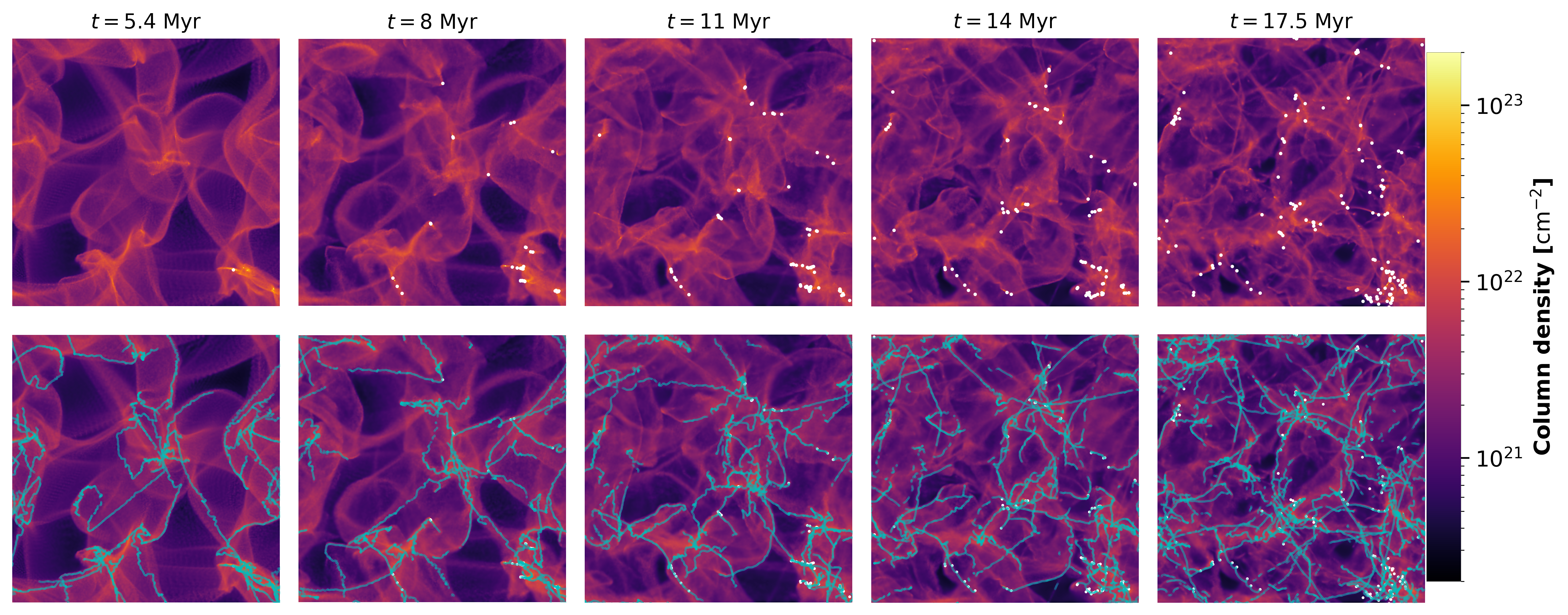}
 \caption{Projected density field in run HDG3 (first row) at times  $t=5.4, 8, 11,14$, and $17.5$ Myr. The second row shows the spines of the filaments in cyan detected by the DisPerSE on top of the density field. White dots represent the sink particles.}
 \label{fig:snaps_sims}
\end{figure*}

We simulate a $256$ pc box with periodic boundaries for hydrodynamics, with a resolution of $145^{3}$. The simulation has an initial uniform density of $n(t=0) = 3\, \pcc$ and a temperature of $T(t=0) = 730$ K, the equilibrium temperature at this density. This corresponds to a mass resolution of $0.5$ M$_{\odot}$. Cooling and heating functions taken from \cite{KI02} were also included, with the typographical correction of \citet{Vazquez-Semadeni+07}. Forcing was applied during the first  $0.65$ Myr for wave numbers such that $1<kL/2\pi<4$, considering a  solenoidal weight of $w (\in [0,1]) = 0.4$, that is, a ratio of compressive to total forcing power of $\approx 0.5$ for 3D data \citep{Federrath+2010a}. After the forcing period, an rms sonic Mach number of $\mathcal{M}_{\rm s}\sim 4.3$ is reached, equivalent to a velocity dispersion of $10.7$ km/s, with a maximum of $\mathcal{M}_{\rm s} \sim 9.1$ at $t=3.4$ Myr, increase due to the formation of cold gas. Shocks are captured using artificial viscosity with a modified version of the switch implemented by \citet{Cullen.Dehnen2010}, which increases dissipation near developing shocks while suppressing it away from shocks. The density for sink formation was fixed at $3.3 \times 10^{6}\, \pcc$, and are capable of accreting and merging. Due to resolution, sinks represent stellar groups rather than individual stars. The critical radius (inside which no new sinks form around an existing one), the accretion radius (where particles are accreted subject to additional criteria), and the merge distance (below which sinks merge unconditionally) were all set to $0.04$ pc, chosen to be identical in order to avoid small time steps during the formation of nearby sinks. As we will be tracking the evolution of sink spin, in Appendix \ref{Appe:AM tratment} we discuss in detail the treatment of the angular momentum of the sinks during the accretion of SPH particles and mergers with other sinks, which considers not only the contributions from the particles and pre-existing sinks, but also the orbital angular momentum. These simulations do not contain any star feedback prescriptions, nor magnetic field. The times of interest are those from the formation of the first sink, at time $t=5.4$ Myr, onwards. The final time considered is $17.5$ Myr. In the first row of Figure \ref{fig:snaps_sims}, we show the projected density field for run HDG3, at times $t=5.4, 8, 11,14$, and $17.5$ Myr. White dots correspond to the sinks particles. 

In the absence of a single, unified nomenclature, we adopt the frequent definition of ``cores'', ``clumps'', and ``clouds'' as structures of sizes $\sim 0.1,1$ and $10$ pc, respectively. Our main analysis focuses on filamentary structures spanning from a few tens of parsecs down to scales of $\sim 0.3$ pc, corresponding to a smoothing length at a density of $\sim 10^3$ cm$^{-3}$, approaching the core regime. However, we also performed two additional local zoom-in simulations, each focused on one filament of particular interest. The zoom-in initial conditions were generated by applying particle splitting to a parent snapshot, replacing each selected SPH particle by 64 equal-mass daughter particles arranged in a centered regular pattern around the parent position. This procedure conserves mass, linear momentum, and orbital angular momentum of each parent particle. The splitting was applied approximately $1$ Myr before the analysis time, shortly before the formation of the corresponding filament (see Appendix \ref{Appe:Fil 1 compare resolution}), and on a volume greater than the volume of interest, allowing us to follow the subsequent filament evolution at higher resolution while giving the daughter-particle distribution time to relax. The resulting particle mass is $m_{\rm part}\sim 0.007$ M$_\odot$. At the new mass resolution, the maximum resolvable density is $n_{\rm res}\sim 4\times10^6 {\rm cm^{-3}}$, corresponding to a smoothing length $h \sim 0.005$ pc. For these simulations, the temporal resolution (time between snapshots) was likewise increased from $0.1$ to $0.01$ Myr.

\section{Methods}
\label{sec:Methods}

\subsection{Filament identification}
\label{subsec:Filament identification}

Filament identification is performed using \textsc{DisPerSE} (Discrete Persistent Structures Extractor; \citet{Sousbie2011}), a topology-based algorithm that applies discrete Morse theory to identify voids, walls, clusters, and filaments from a scalar field defined in space. In this work, the input to \textsc{DisPerSE} is an interpolated density field mapped onto a 3D regular grid. This mesh is constructed from the \textsc{Phantom} simulation outputs using the \textsc{Splash} visualization and analysis tool \citep{Price2007}, and the Python package \textsc{Sarracen} \citep{Harris.Tricco2023}, which provides native support for reading binary outputs in Python.

The \textsc{DisPerSE} algorithm incorporates a persistence threshold that sets a lower limit for identifying structures that are statistically significant with respect to the background noise. For the analysis of filaments across the entire computational domain, a persistence threshold corresponding to a density of $\sim 2 \times 10^{2}$ cm$^{-3}$  was adopted. In addition, selected filaments are analyzed individually, for which the persistence threshold is adjusted to better isolate the structure of interest. The second row of Figure \ref{fig:snaps_sims} displays the filaments identified by \textsc{DisPerSE}, shown in cyan overlaid on the density field. A rich network of filamentary structures is recovered, becoming increasingly complex as the simulation evolves.

\subsection{Core angular momentum approximation}
\label{subsec:outflow approx}

\begin{figure*}
\centering \offinterlineskip
\includegraphics[width=\linewidth]{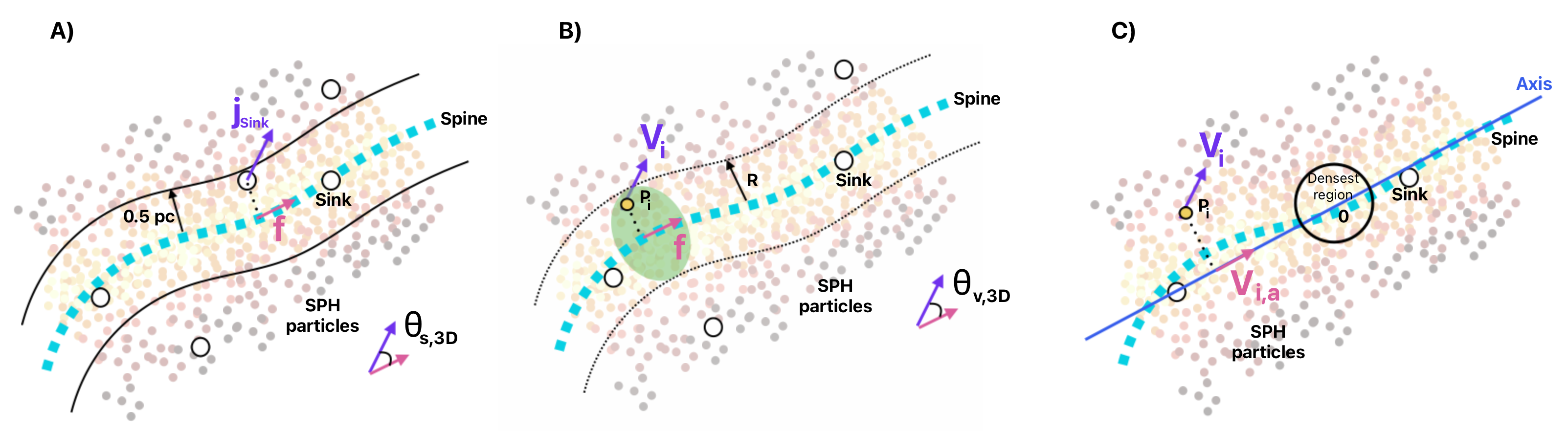}
\caption{\textit{A)} Representative scheme for the calculation between the angular momentum vector of a sink (large white circles) represented in purple, ${\bm j}_{\rm sink}$, and the filament direction shown in pink, calculated from the point on the spine closest to the sink. Sinks are assigned to a filament if they are at a distance $\leq 0.5$ pc from the spine. \textit{B)} Schematic drawing showing the calculation of the angle between the bulk-subtracted velocity vector of the particles around the filament, ${\bm V}_i$, and the filament direction, \textbf{f}. A test SPH particle, $P_i$, is shown as an example in yellow. The filament direction is also calculated from the point on the spine closest to the particle. The process is repeated for all particles around the spine for $R=0.5$ pc. \textit{C)} Projection of the SPH particle velocity vectors along the major axis of the filament, determined from a 3D fit of the spine points. The zero position is defined as the projected spine point whose associated particles have the highest mean density.}
 \label{fig:angle calculation}
\end{figure*}

As mentioned in the previous section, no prescription for stellar feedback is included in this simulation, consequently, the formation of actual protostellar outflows cannot be followed explicitly. Given the resolution of our simulations, we adopt the angular momentum of the sinks as a proxy for the angular momentum of the corresponding  cores, considered as unresolved fragments. Although the final orientation of a protostellar disk is determined by the local dynamics of the collapsing envelope and disk system, as shown by \citet{Tsukamoto.Machida2013}, the angular momentum supplied to the cores can still influence the initial conditions for this subsequent evolution. In the present simulations, longitudinal flows along filaments provide continued accretion onto the cores, and therefore also a sustained source of angular momentum. Thus, we do not regard the sink angular momentum as a direct prediction of the final disk or outflow orientation, but rather as a measure of the large-scale angular momentum delivered to the collapsing system before smaller-scale processes take over.

Once a sink forms, the closest point along the filament spine identified by \textsc{DisPerSE} is found, and the sink is associated with that filament if its distance to the spine is $\leq 0.5$ pc. The local filament direction is then defined using the nearest consecutive spine points. Finally, we calculate the minimum 3D angle, $\theta_{\rm 3D} \in [0\degree,90\degree]$, between sink's angular momentum vector and filament's direction. In the left panel of Figure \ref{fig:angle calculation} we show a representative scheme for calculating the angle between the angular momentum vector of the sinks (large white circles) in purple (${\bm j}_{\rm sink}$), and the direction of the filament denoted in pink (\textbf{f}). We repeat this procedure at each time step, following the sinks from their formation to the final time studied ($t=17.5$ Myr), or until they merge with a more massive sink, in which case they are considered as inactive or dead. We also defined a second subsample of sinks considering their contribution to the angles distribution during the first $0.5$ Myr from the time of their formation.

The full sink sample, that is, with not time restriction, is used to investigate whether a transition in the orientation of angular momentum occurs as gravity becomes increasingly important, thereby testing the alignment mechanism proposed by \citet{Anathprindika.Withworth2008}. In contrast, the restricted subsample is designed to assess whether angular momentum vectors exhibit a preferential orientation at birth, depending on the evolutionary stage of the simulation, or whether such orientations emerge later as a consequence of gas dynamics. In other words, this approach allows us to distinguish between angular momentum orientations that are primordial and those that result from subsequent dynamical reorientation.

\subsection{Velocity field around filaments}
\label{subsec:velfield in filaments}

The velocity field around the filaments can be examined to assess whether gas motions exhibit a preferential orientation with respect to the filament axis. To this end, the 3D angle between the velocity vector of SPH particles in the vicinity of the filaments and the local filament direction, $\theta_{\rm v,3D}$, is computed to determine the presence of any preferred alignment. Accretion motions onto the filament are expected to produce predominantly perpendicular orientations between these vectors, whereas longitudinal flows along the filament axis should manifest as a more parallel alignment.

To measure $\theta_{\rm v,3D}$, curved cylindrical volumes of radius $R = 0.5$ pc are defined around the filament spines identified with \textsc{DisPerSE}. All SPH particles within this radius are associated with their nearest spine point. To remove contributions from bulk motions, the mean velocity of the particles assigned to each spine point is subtracted from the velocity of each individual SPH particle. The local filament direction at each spine point is computed following the same procedure described in the previous section for the determination of $\theta_{\rm s,3D}$. The minimum 3D angle between the particle velocity vector and the filament direction is then calculated. A schematic illustration of this procedure is shown in panel \textit{B} of Figure \ref{fig:angle calculation}, here the velocity vector, (${\bm{\mathrm{V}}}_{i}$), is indicated in purple, and the filament direction by a pink vector (\textbf{f}), for a representative particle, $P_{i}$, shown in yellow. This procedure is repeated for all spine points in all filaments identified within the simulation domain.

The two zoom-in filaments introduced in Section \ref{sec:Numerical data} were also analyzed individually to assess whether the presence or absence of longitudinal flows is related to the emergence of a preferential alignment in the $\theta_{\rm s,3D}$ distribution. $\theta_{\rm s,3D}$ was computed following the same procedure described in the previous section for the full filament sample. In addition, to characterize the velocity field and evaluate the role of longitudinal flows, a single three-dimensional principal axis was fitted to the spatial distribution of points along the filament spine. The velocity vectors of the SPH particles were then projected onto this axis to construct a velocity profile along the filament’s major axis. The filament center, defined as the zero point in both position and velocity, was taken to be the projected spine point position whose associated SPH particles exhibit the highest mean density. A schematic illustration of the velocity decomposition is shown in panel \textit{C} of Figure \ref{fig:angle calculation}, where the velocity of an SPH particle (${\bm{\mathrm{V}}}_{i}$) is projected onto the filament’s major axis (solid dark blue line) and obtaining (${\bm{\mathrm{V}}}_{i,{\rm a}}$).

By adopting this axis definition for individual filaments, we aim to capture the overall tendency of the velocity field driven by gravity, which does not necessarily follow the exact spine geometry. Furthermore, defining a single principal axis is only feasible for isolated filaments that are sufficiently straight to admit a well-defined direction of elongation. This approach is not applicable when considering the full filament population, as these structures are typically curved and interconnected within a complex network. In such cases, the filament orientation was determined locally by estimating the tangent direction at each reference spine point using its immediate neighboring points, as described above.

\section{Results}
\label{sec:Results}

\subsection{Full-box analysis}
\label{subsec:Full-box analysis}

\subsubsection{Sink angular momentum-filament alignment}
\label{subsec:OFA in full-box}

\begin{figure}
\centering \offinterlineskip
\includegraphics[width=\linewidth]{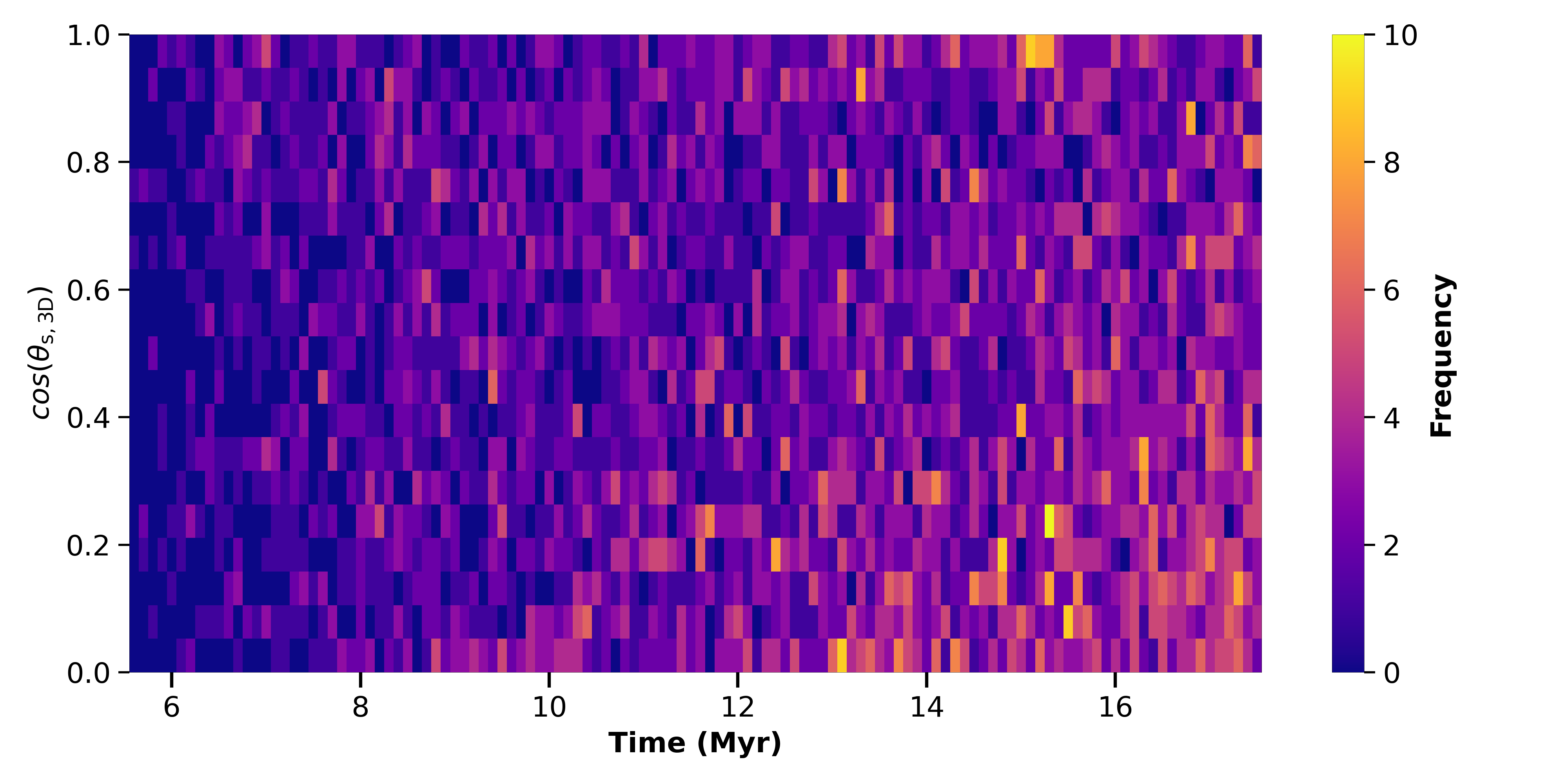}
\includegraphics[width=\linewidth]{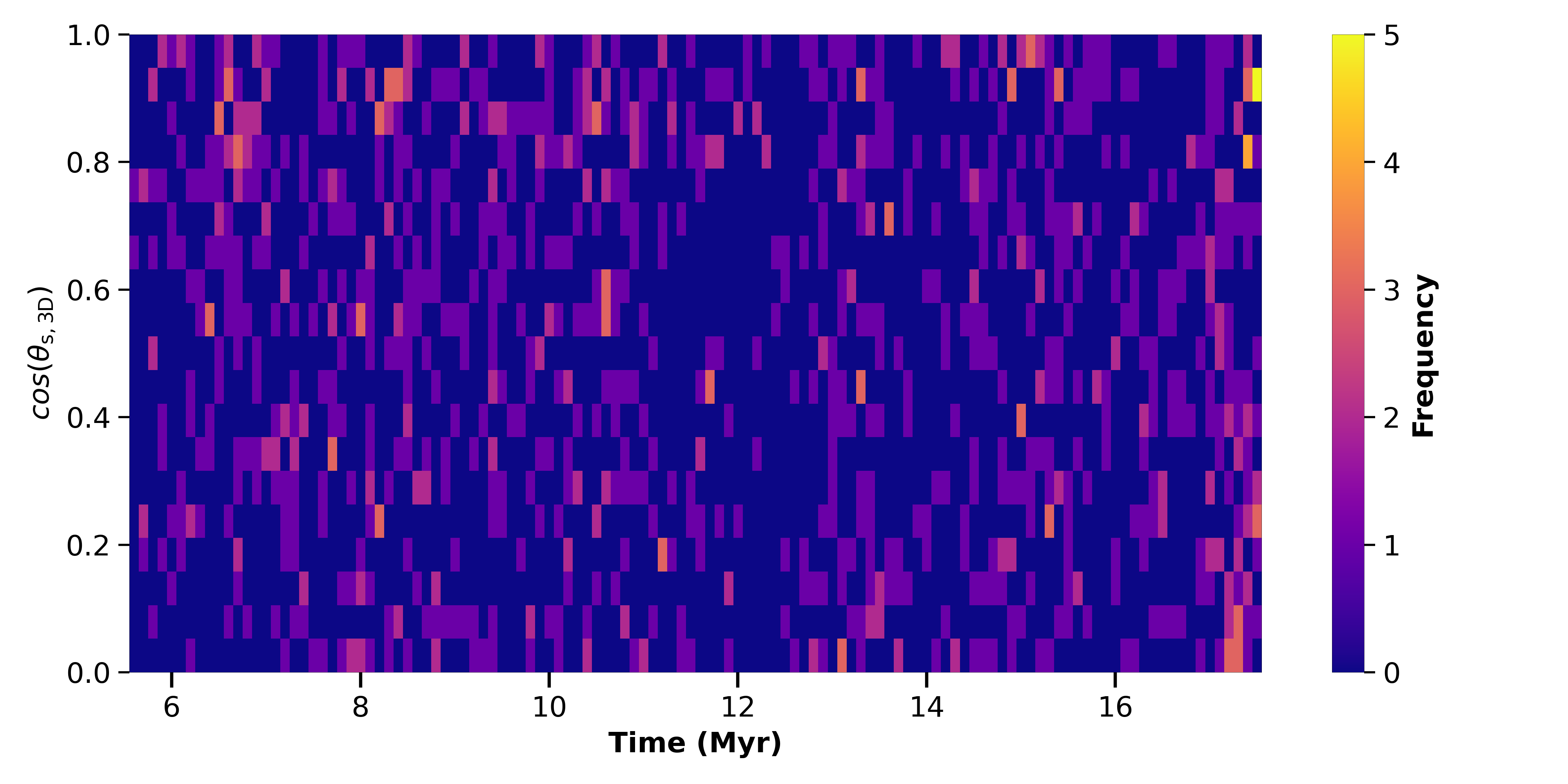}
 \caption{{\it Top panel:} Evolution of the histograms of $cos(\theta_{{\rm s,3D}})$ for the angle between the angular momentum vector of the sinks and the direction of the associated filament for the full numerical sample of sinks. {\it Bottom panel:} Same as top panel but the sinks are tracked only during the first $0.5$ Myr after their formation. A tendency towards perpendicularity can be seen at later times for the sinks without the time restriction, while in the reduced sample, the overall trend seems rather random.}
 \label{fig:OFA evolution with gravity}
\end{figure}

It is well known that, unlike projected 2D angles, the distribution of 3D angles between pairs of randomly generated 3D vectors is not uniform, rather, it peaks at $90\degree$ \citep{Stephens+2017}. To avoid statistic artifacts, in Figure \ref{fig:OFA evolution with gravity}, we show the evolution of the histograms of $cos(\theta_{{\rm s,3D}})$. Each histogram is plotted vertically at each time step, with the color coding representing the frequency per bin. Top panel shows this evolution for all active sinks within the simulation domain over the time period considered. It can be seen that at early times ($t < \sim10$ Myr), the distribution appears random, while at later times a tendency towards perpendicularity emerges, with high frequencies (light-colored bins) in the bottom right quadrant. Therefore, we consider this to be a trend resulting from physical processes and not due to geometric effects. 

This behavior can be explained either by the reorientation of the angular momentum of the sinks at later times in the simulation, or by new sinks formed with an initial preferred orientation. To further elucidate this point, we conducted a second tracking considering sinks during the first $0.5$ Myr after their formation, as described in Section \ref{subsec:outflow approx}. In the bottom panel of Figure \ref{fig:OFA evolution with gravity}, we show the evolution of $cos(\theta_{{\rm s,3D}})$ for this sample. Compared with the non-restricted sampled in the top panel, no clear tendency is found. The overall trend seems rather random. This suggests that the excess of large angles (i.e., near-perpendicular alignments) observed at later times is primarily the result of a subsequent reorientation of sinks that formed earlier, leading to an increase in $\theta_{{\rm s,3D}}$ toward perpendicular configurations over time. To illustrate this, Figure  \ref{fig:trayectory over OFA evolution} in Appendix \ref{Appe:Sink trajectories} shows the evolution of the angles between the angular momentum vector and the host filament for three sinks selected as those that survive the longest.

\begin{figure}
\centering \offinterlineskip
\includegraphics[width=\linewidth]{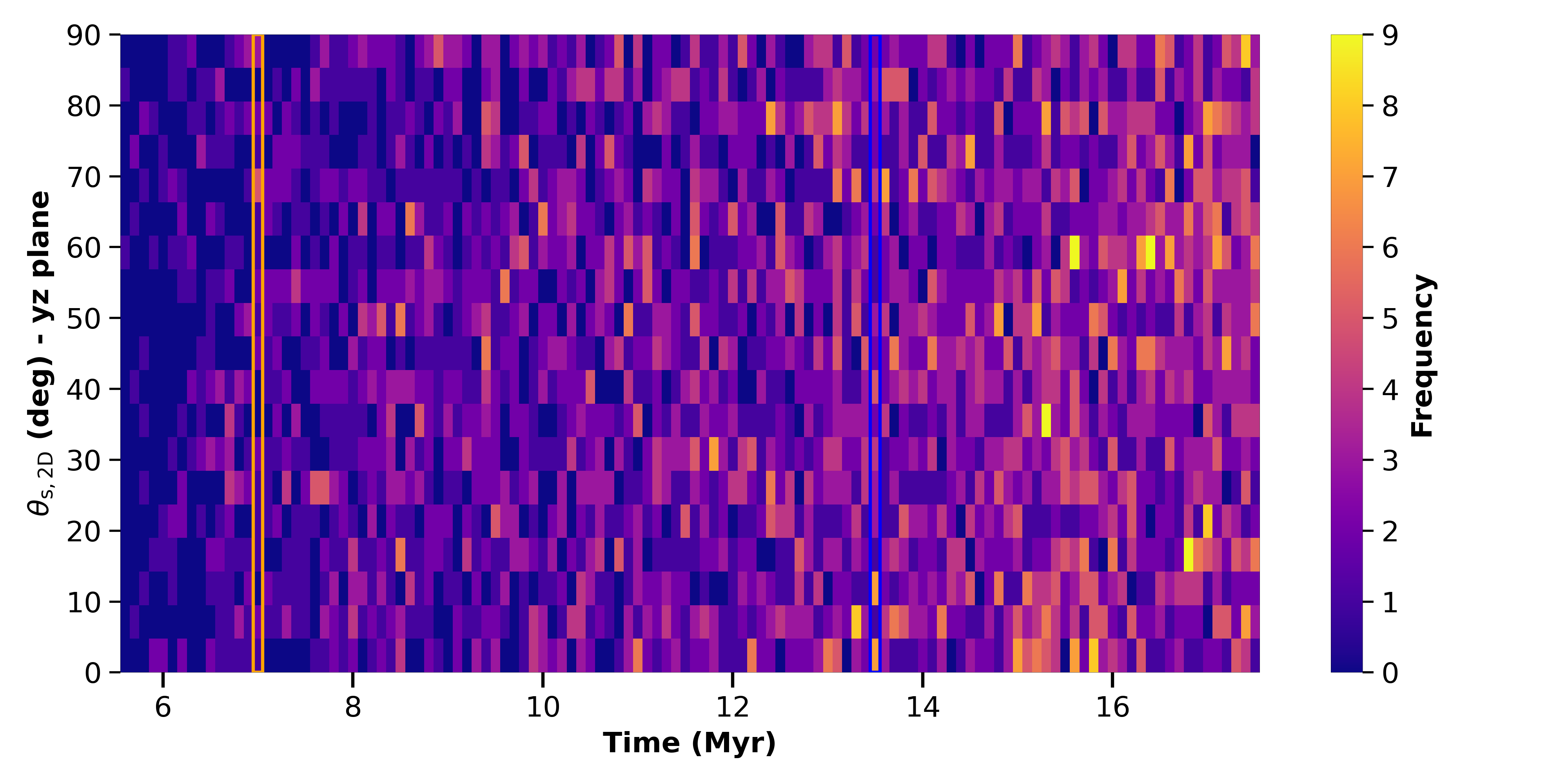}
\includegraphics[width=\linewidth]{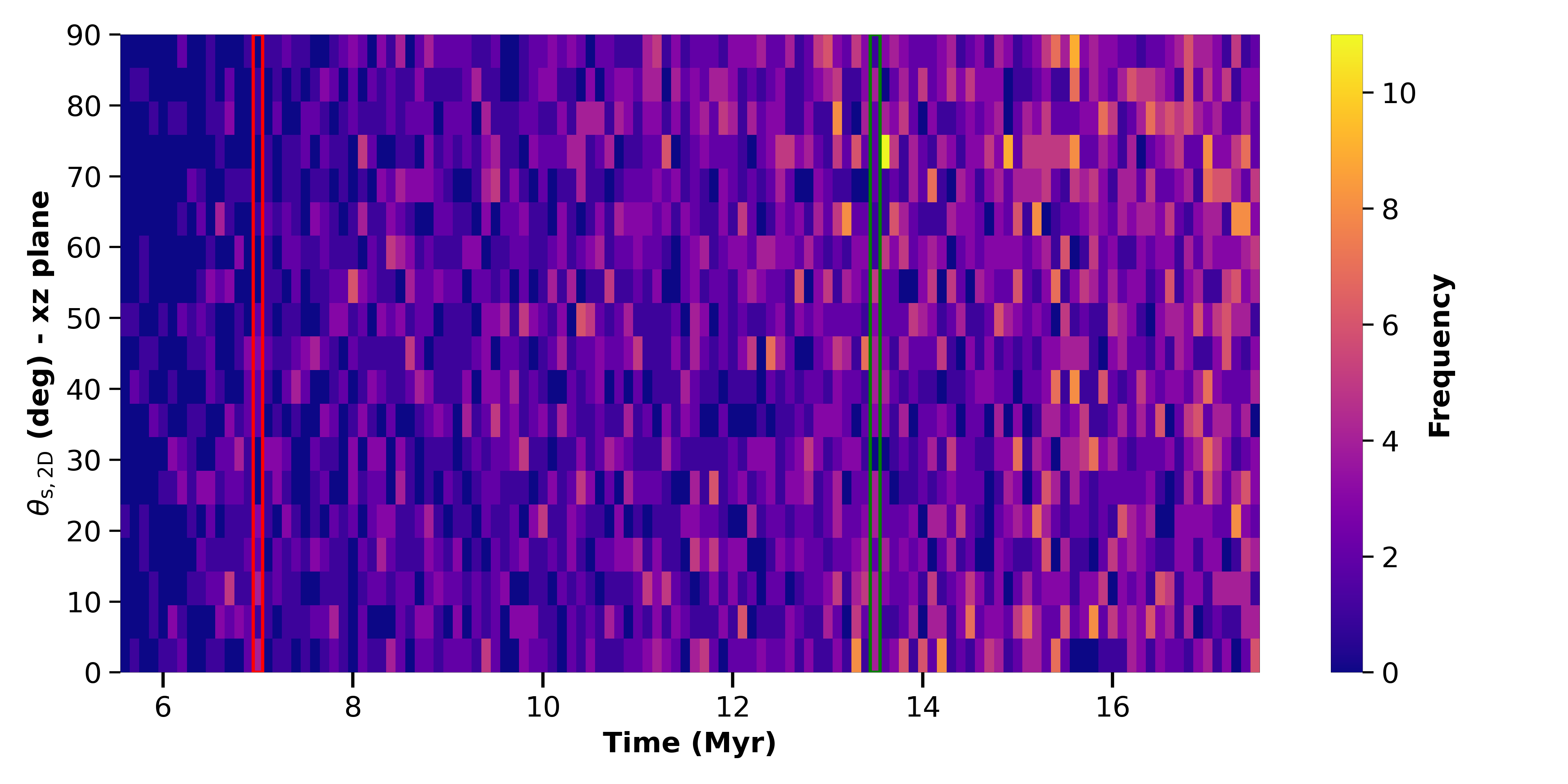}
\includegraphics[width=\linewidth]{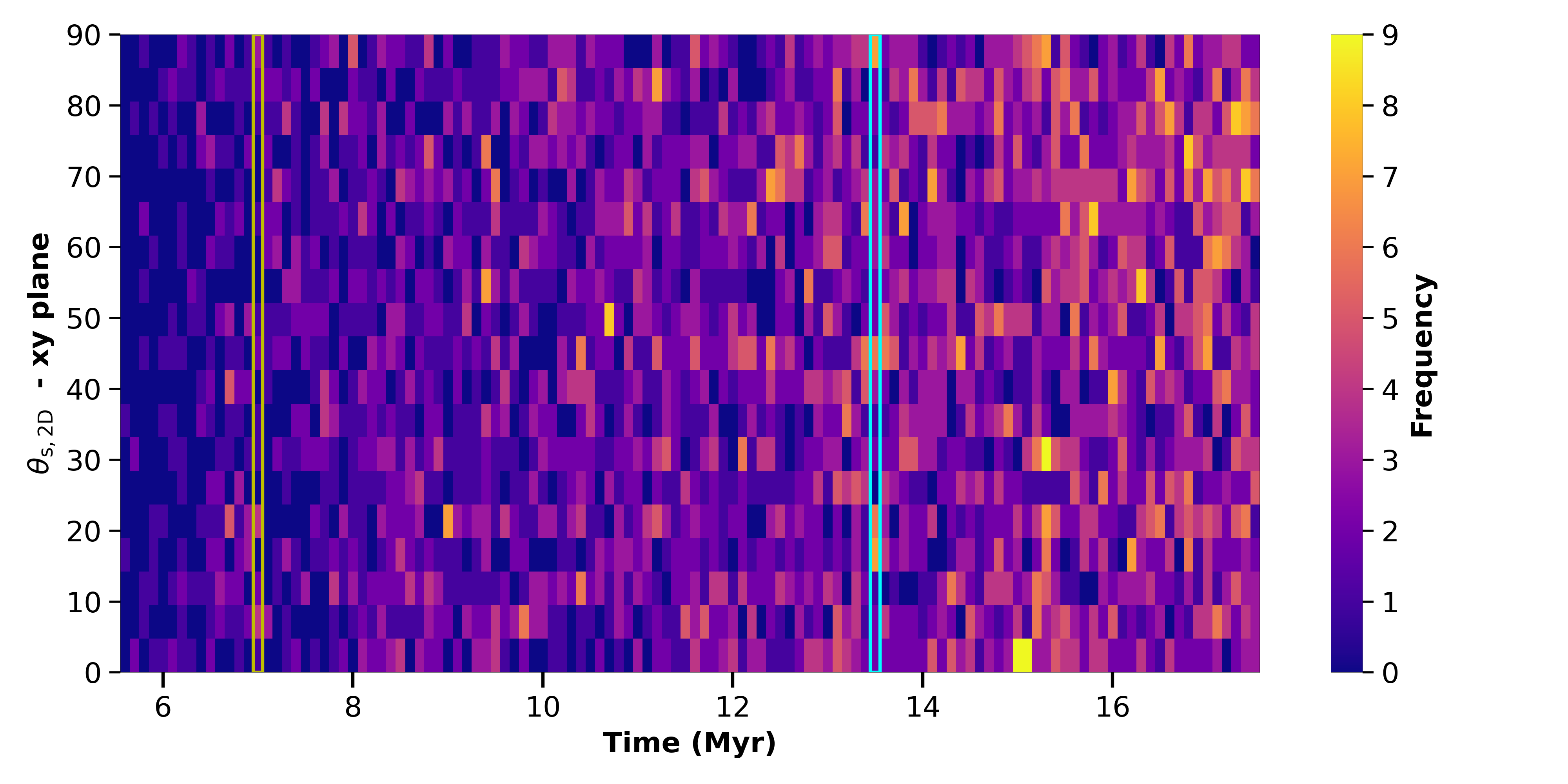}
 \caption{Evolution of $\theta_{{\rm s,2D}}$, measured as the angle between the 2D projections of the angular momentum vectors of the sinks and the filament orientation onto the three coordinate planes, for the full sample with no lifetime restriction. Early (warm colors) and late (cool colors) times highlighted with vertical lines, are shown individually as cumulative distributions in Figure \ref{fig:2D OFA examples} using the same color code. No particular trend over time is observed in any of the projections, unlike its 3D counterpart.}
 \label{fig:2D OFA evolution}
\end{figure}

\begin{figure}
\centering \offinterlineskip
\includegraphics[width=\linewidth]{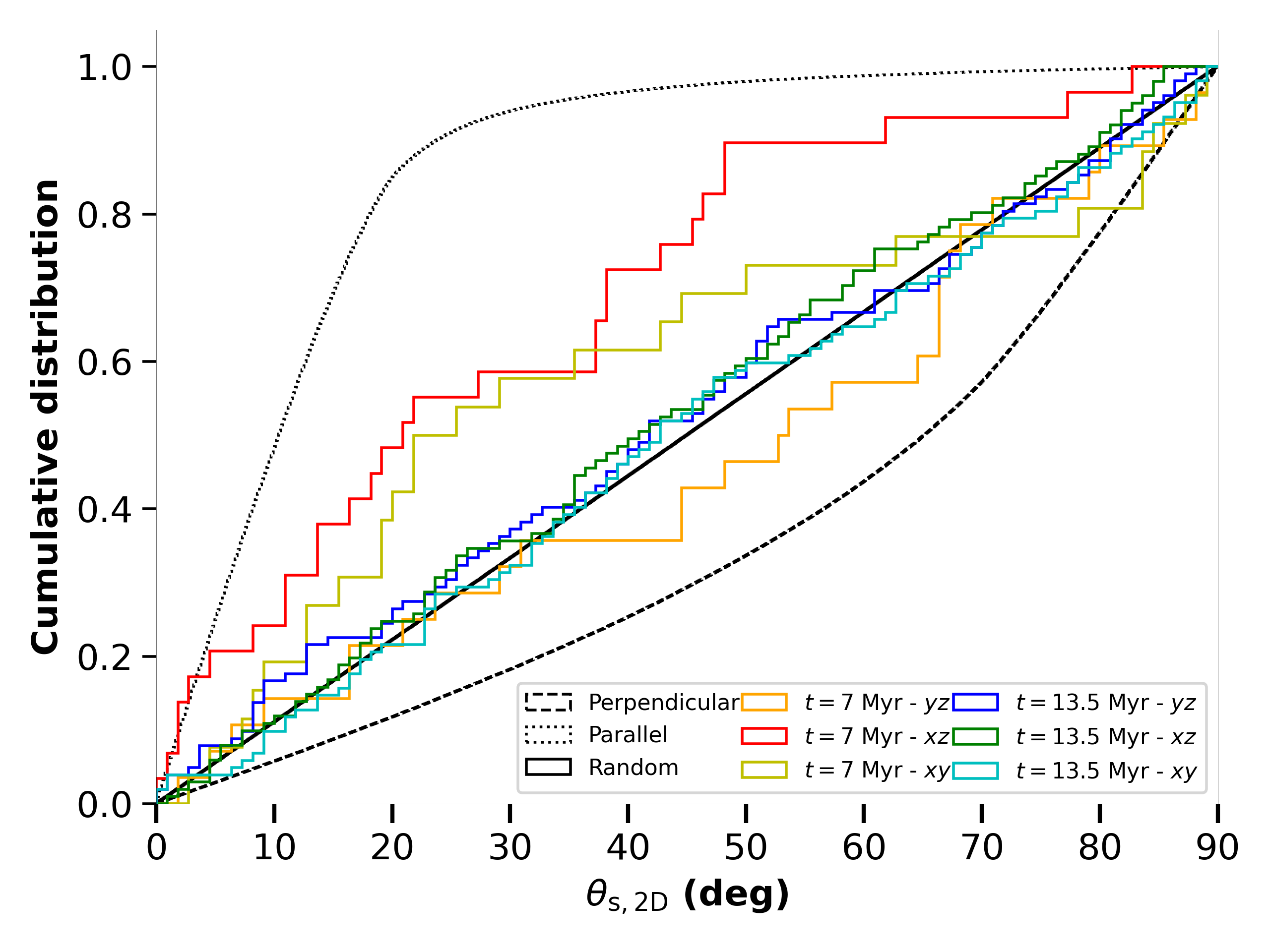}
 \caption{Cumulative histograms of the angle between the 2D projection of the angular momentum vector of the sinks and the filament orientation, $\theta_{{\rm s,2D}}$, at times $t=7$ (warm colors) and $13.5$ (cold colors) Myr (highlighted in Figure \ref{fig:2D OFA evolution}) on the three coordinate planes $xy$, $xz$, and $yz$. No lifetime restrictions were applied. At time $t=7$ Myr the distribution varies greatly depending on the projection, while at time $t=13.5$ Myr, it appears random in all three projections.}
 \label{fig:2D OFA examples}
\end{figure}

In order to compare with observations, which only have access to information on the plane of the sky, we perform 2D projections of the angular momentum vectors of the sinks and the filament orientation onto the three coordinate planes, and measure the angle between them, $\theta_{{\rm s,2D}}$. We seek to determine whether the trend towards perpendicularity observed in 3D also occurs in 2D projections. Figure \ref{fig:2D OFA evolution} shows the evolution of the $\theta_{{\rm s,2D}}$ histograms projected onto the three coordinate planes. Unlike the three-dimensional case, no trend is observed in any of these planes, indicating that the projected trend is a random distribution. However, it is possible to identify times for which the distribution under a certain projection may appear moderately parallel or perpendicular, even for the same time. In Figure \ref{fig:2D OFA examples} we take two snapshots at times $t=7$ (warm colors) and $13.5$ (cold colors) Myr (highlighted in Figure \ref{fig:2D OFA evolution}), and we plot the cumulative histogram of $\theta_{{\rm s,2D}}$ for the three projections at each time. Reference distributions corresponding to random (solid line, $0\degree - 90\degree$), preferentially perpendicular (dashed line, $70\degree - 90\degree$), and preferentially parallel (dotted line, $0\degree - 20\degree$) orientations are shown for comparison, following the convention commonly adopted in theoretical/observational studies of outflow-filament alignments. The histograms vary greatly at $t=7$ Myr, showing moderately perpendicular alignment on the $yz$ plane, while in the $yz$ and $xz$ planes it appears to lean toward a moderately parallel orientation. On the other hand, at time $t=13.5$ Myr, the distribution appears to be mainly random in all three projections. This appears to be the dominant case at late times in the simulation, where as more sinks are formed, the distribution appears to plateau at a random distribution. This suggests that, as might be expected, 2D projections reveal a fraction of the complete 3D information. 

\subsubsection{Velocity vector-filament alignment}
\label{subsubsec:velfield in ful-box}

\begin{figure}
\centering \offinterlineskip
\includegraphics[width=\linewidth]{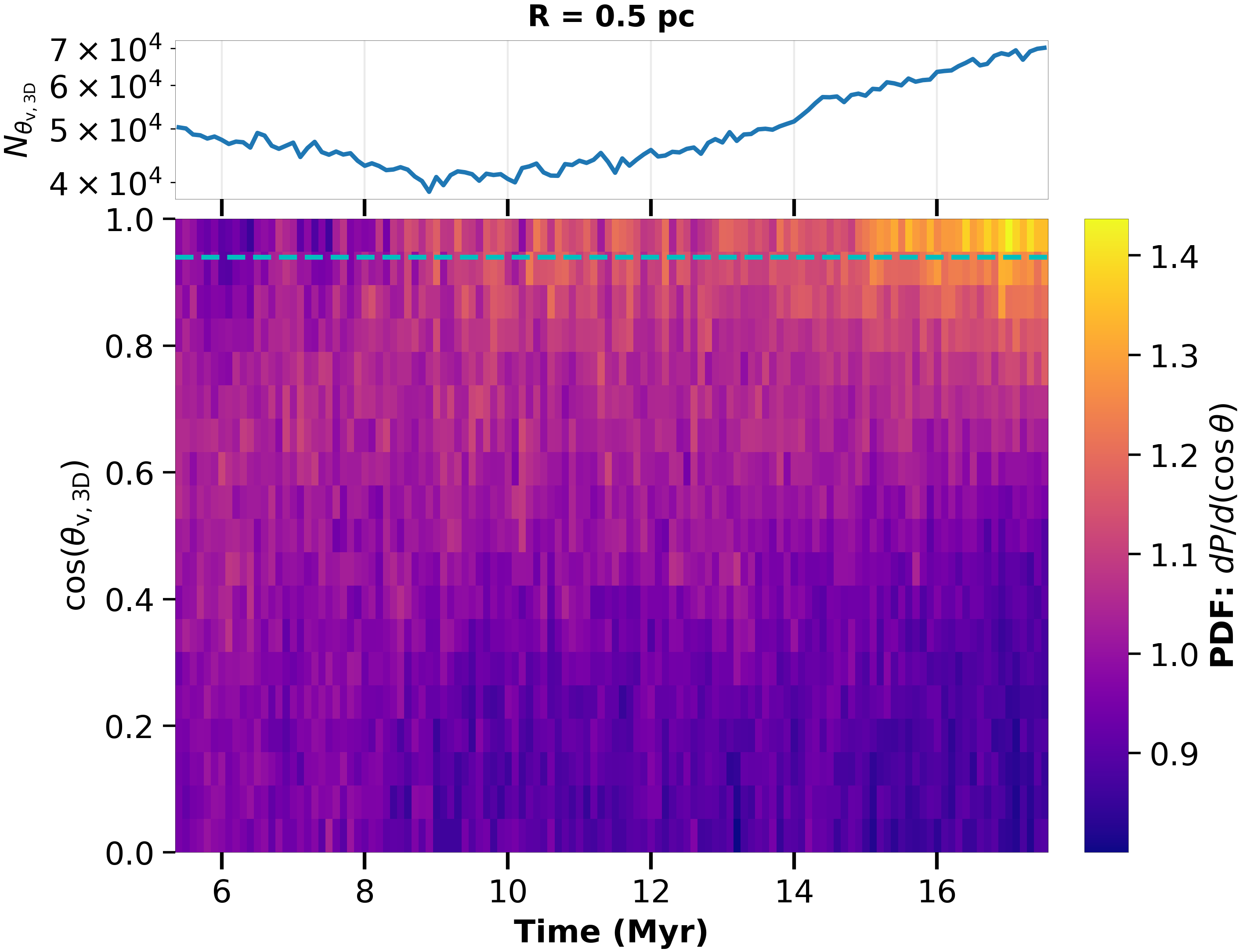}
 \caption{Evolution of the probability distribution function of $cos(\theta_{{\rm v,3D}})$, the cosine of the 3D angle between the velocity vectors of SPH particles located within $0.5$ pc of the filament spine and the local filament orientation (see panel \textit{B} of Figure \ref{fig:angle calculation}). Dashed cyan line marks $\theta_{\rm v,3D}=20\degree$. An increase in parallel alignments (upper right corner), corresponding to angles close to zero ($cos(\theta_{{\rm v,3D}})\sim 1$), can be seen at later times in the simulation, which may indicate the development of longitudinal flows along filaments.}
 \label{fig:Velocity field-filament alignment}
\end{figure}

To characterize the gas dynamics within the filaments in a global manner, we computed the time evolution of the 3D angle, $\theta_{\rm v,3D}$, between the velocity vectors of SPH particles located within $0.5$ pc of the filament spine, and the local filament orientation, as illustrated in panel \textit{B} of Figure \ref{fig:angle calculation}. Figure \ref{fig:Velocity field-filament alignment} shows the evolution of the probability distribution function of $cos(\theta_{{\rm v,3D}})$, and the dashed cyan line marks $\theta_{\rm v,3D}=20\degree$. In the early stages, a deficit of angles near $0\degree$ ($\cos(\theta_{\rm v,3D})\simeq 1$) is observed, consistent with accretion flows directed toward the filament predominantly perpendicular to its major axis. As the simulation evolves, this deficit gradually shifts toward configurations clustered around $0\degree$. At later times in the simulation, when gravity is expected to dominate the dynamics, a clear excess of angles close to zero ($\cos(\theta_{\rm v,3D})\simeq 1$; upper-right corner) emerges. This behavior appears to reveal the formation of longitudinal flows parallel to the filament axis and is consistent with the contemporaneous appearance of angles close to $90^\circ$ in the distribution of angles between the sinks’ angular momentum vectors and the local filament direction. The top panel presents the evolution of the total number of angles ($N_{\theta_{{\rm v,3D}}}$), and therefore, of the SPH particles associated with filaments. $N_{\theta_{{\rm v,3D}}}$ may initially decrease as new sinks form and begin to accrete SPH particles, potentially reducing the number of SPH particles remaining within filamentary structures. This trend could also be associated with the possibility that, following the initial turbulent compressions, some overdense regions do not continue toward sustained gravitational collapse and instead disperse. The subsequent rise in $N_{\theta_{{\rm v,3D}}}$ after $t\sim9$ Myr may reflect a renewed phase of accretion onto the filaments as they grow in mass.

\subsection{Individual filaments}
\label{subsec:OFA in sim with gravity}

\subsubsection{Filament 1}
\label{subsubsec:Filament 1}

\begin{figure*}
\centering \offinterlineskip
\includegraphics[width=\linewidth]{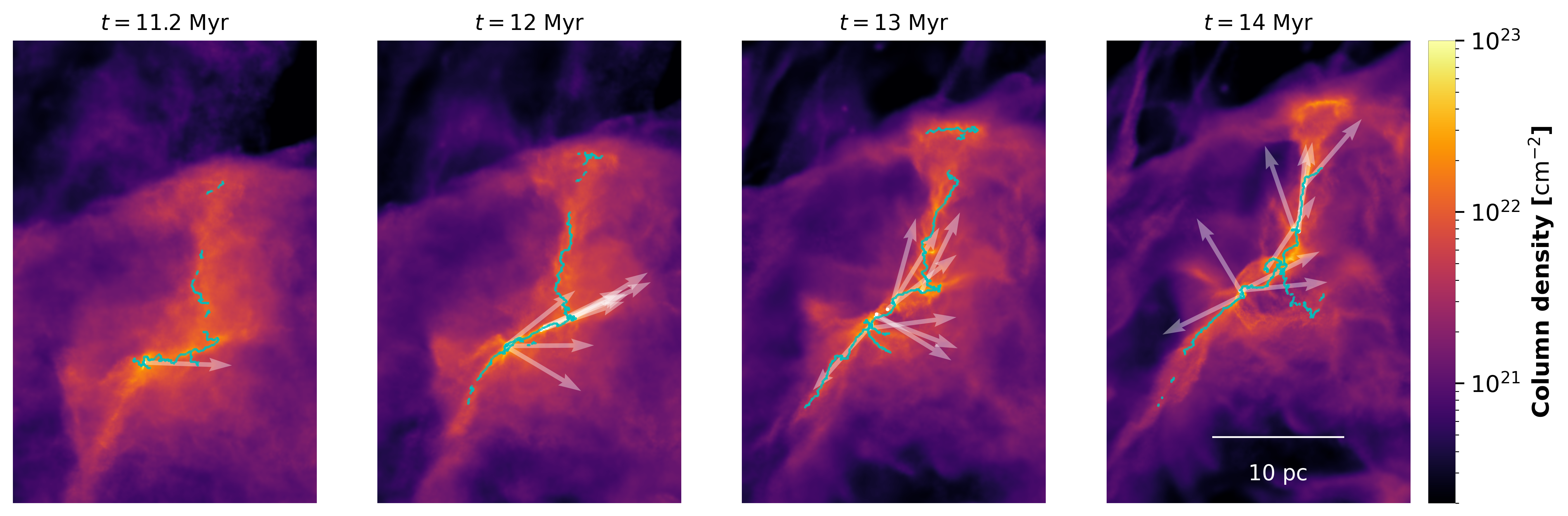}
\includegraphics[width=\linewidth]
{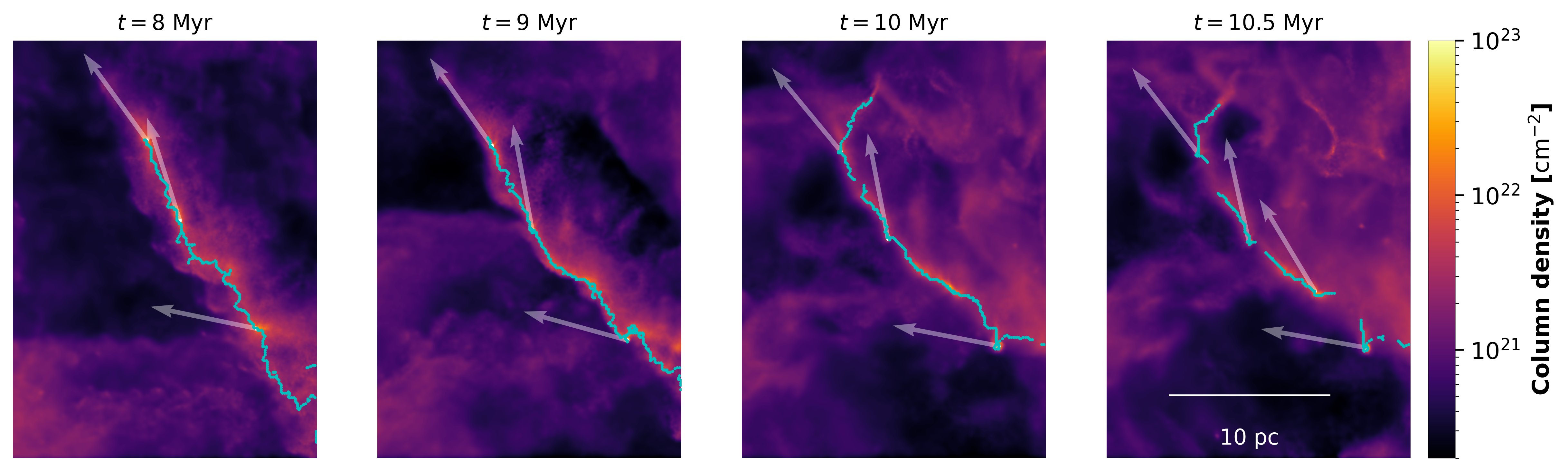}
 \caption{Projected density field for Filament 1 (top panel) at  $t=11.2,12,13$ and $14$ Myr and Filament 2 (bottom panel) at  $t=8,9,10$ and $10.5$ Myr. For each filament, the projected density is shown as well as the spines of the filaments in cyan detected by the DisPerSE on top of the density field. Sink particles are represented by white dots, and the transparent white arrows represent the direction of their angular momentum vector.}
 \label{fig:column density for individual filaments}
\end{figure*}

\begin{figure}
\centering \offinterlineskip
\includegraphics[width=\linewidth]{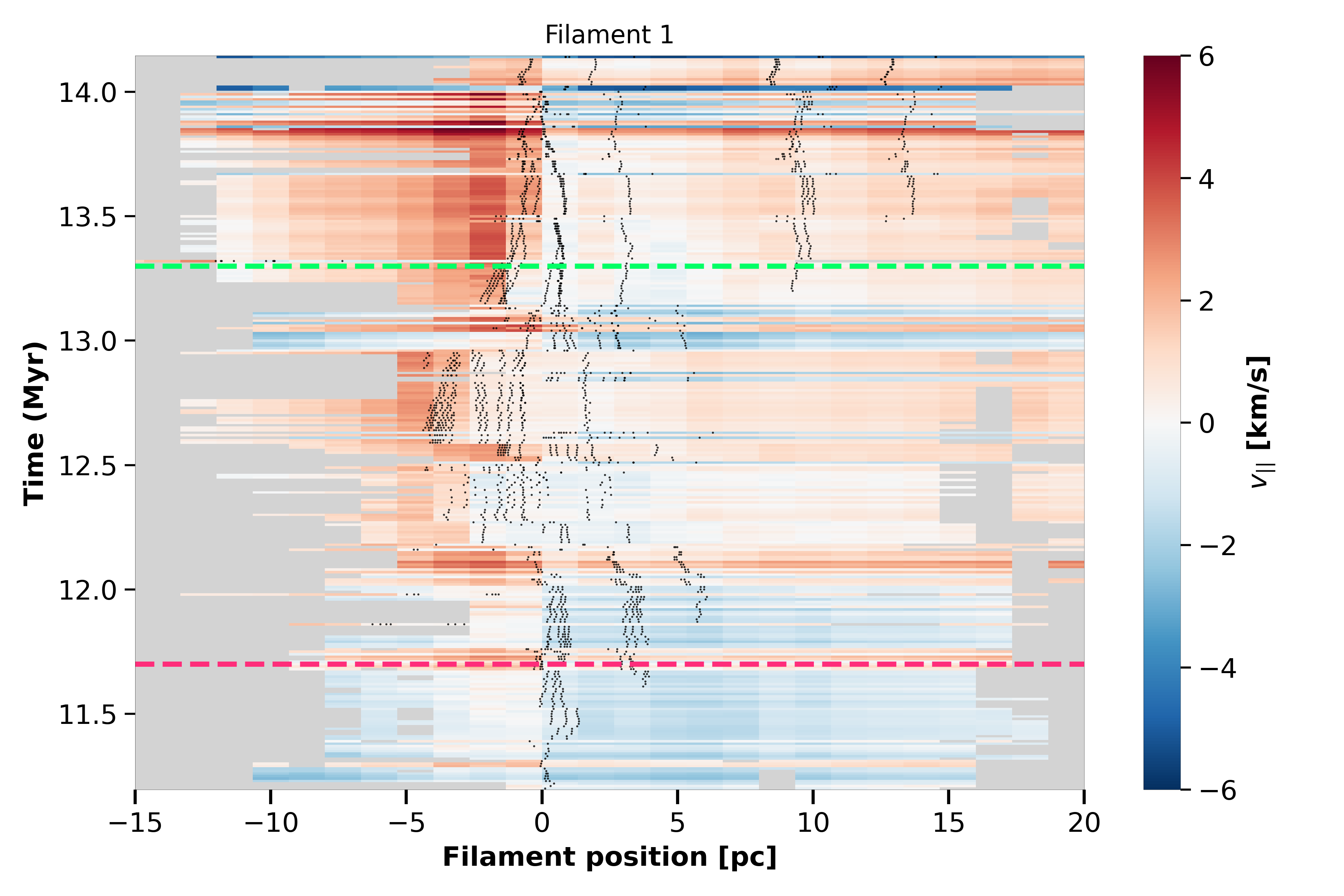}
\includegraphics[width=\linewidth]{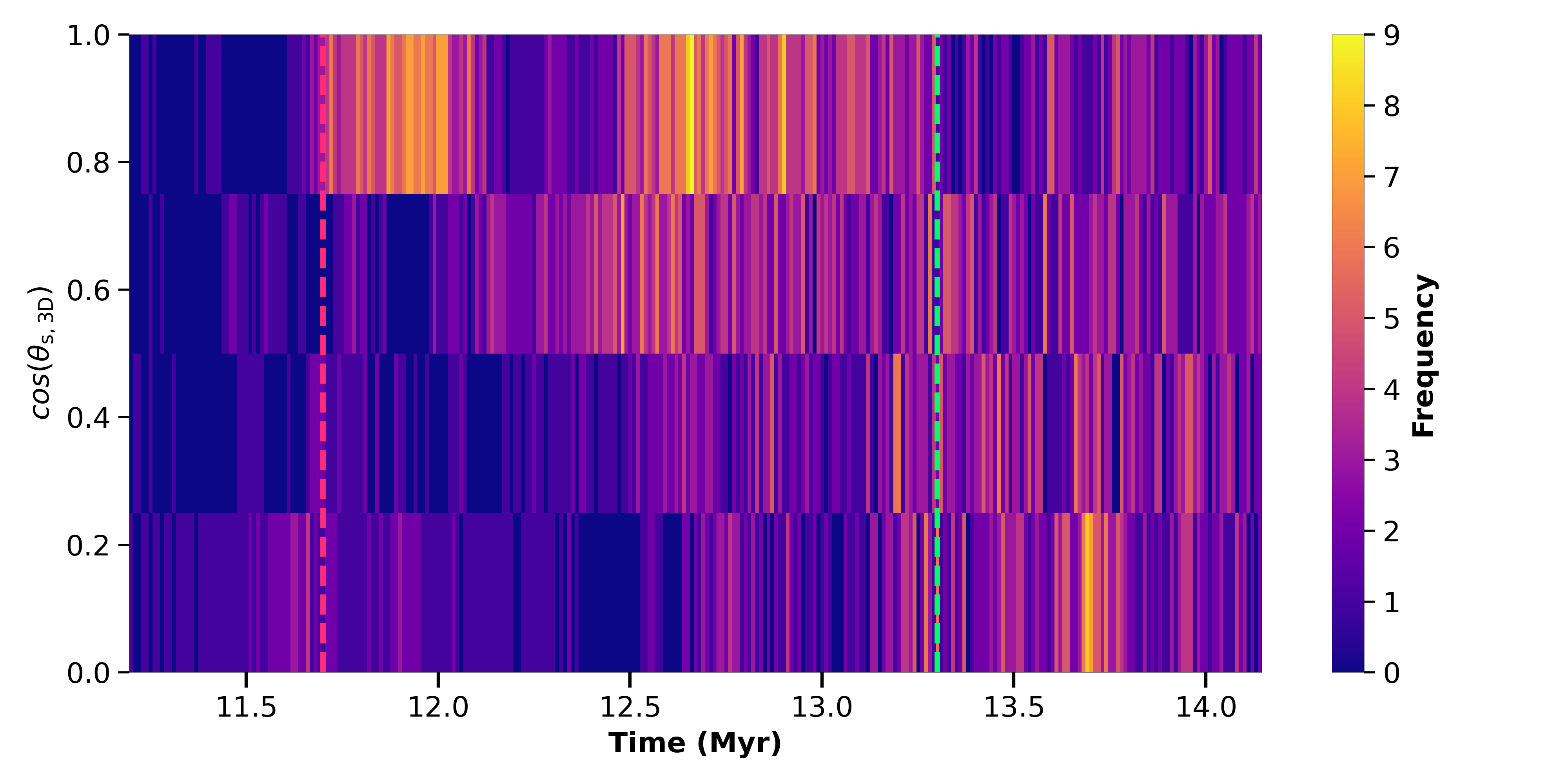}
\includegraphics[width=\linewidth]{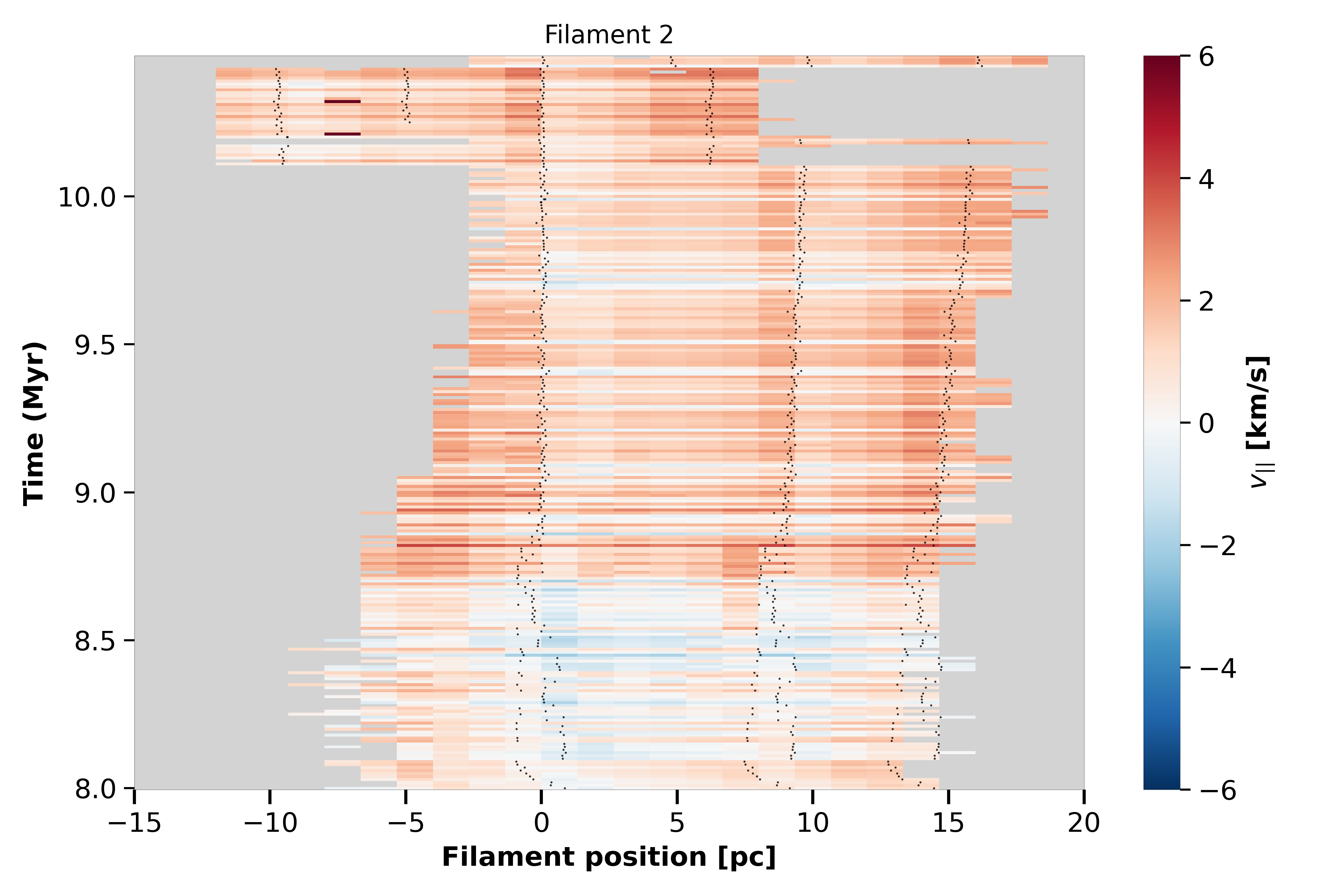}
\includegraphics[width=\linewidth]{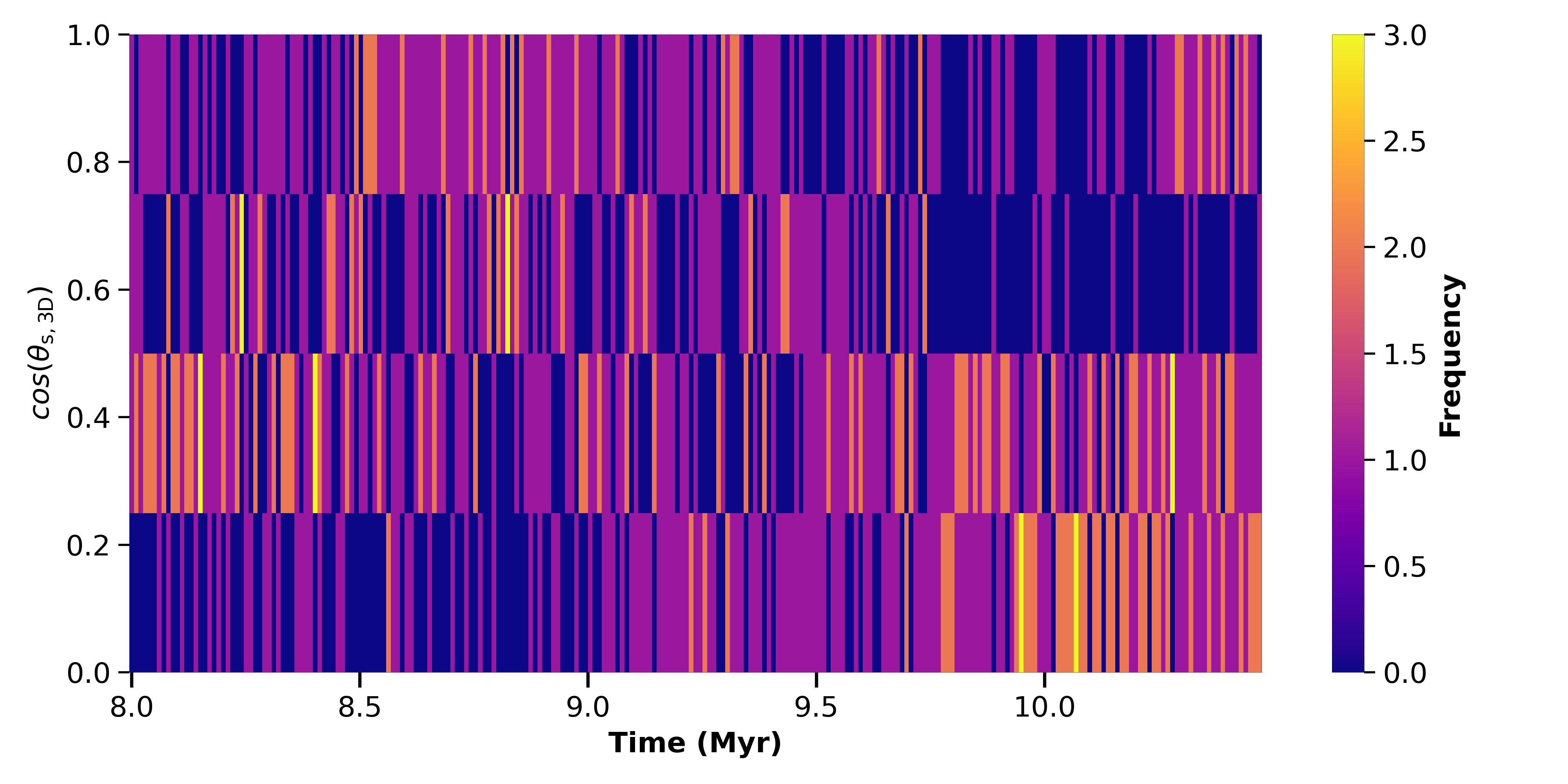}
 \caption{{\it First panel}: Evolution of the velocity projected along the principal axis $v_{||}$ fitted to the spine points of Filament 1 (top row of Fig. \ref{fig:column density for individual filaments}), with the density peak adopted as the zero point of both position and velocity. Black dots represent the projected position of the sinks at each time step. {\it Second panel}: Evolution of $cos(\theta_{{\rm s,3D}})$ histograms for Filament 1 from time $t=11.2$ Myr onwards, with two sinks already formed. No time constraint was applied. {\it Third and forth panel}: Same plots for Filament 2 (bottom row of Fig. \ref{fig:column density for individual filaments})}.
 \label{fig:OFA 3D and flows}
\end{figure}

In the top rows of Figure \ref{fig:column density for individual filaments} we show the projected density field of one single filament at times $t=11.2,12,13$ and $14$ , which we identify as Filament 1. Points belonging to the filament spine provided by \textsc{DisPerSE} are shown in cyan on top of the density field. Sink particles are represented by white dots, and the transparent white arrows represent the direction of their angular momentum vector. A comparison with the low-resolution parent calculation at the same physical time shows that the large-scale filament morphology is preserved, while the refined simulations reveal additional internal structure on smaller scales (See Appendix \ref{Appe:Fil 1 compare resolution}).

We measured $\theta_{\rm s,3D}$ following the same procedure used for the full filament sample. In the high-resolution zoom-in simulations, however, additional small-scale structure develops around the sinks, including disk-like features that can lead \textsc{DisPerSE} to identify a dense concentration of spine points in the immediate sink vicinity. We therefore excluded spine points within a $0.5$ pc radius of each sink when searching for the nearest point on the spine.

For individual filaments no time constraint was applied, so as not to further decrease the already small sample of sinks. In Figure \ref{fig:OFA 3D and flows} we show the evolution of the histograms of $cos(\theta_{{\rm s,3D}})$ for Filament 1 (second row). During the first half-megayear of evolution no particular trend is observed. However, between $t\sim 11.7$ and $t\sim 13.3$ Myr, a tendency to have preferably parallel orientations seems to emerge. From $t\sim 13.3$ Myr onwards, this orientation seems to reorient to a predominantly perpendicular one, similar to the behavior of the sinks presented in Figure \ref{fig:trayectory over OFA evolution} distributed throughout the numerical box.

To assess whether this reorientation is driven by longitudinal flows along the filament, a principal axis was fitted to the spine-point distribution, and the SPH particle velocities were projected onto this axis, as described in Section \ref{subsec:velfield in filaments} and shown in panel \textit{C} of Figure \ref{fig:angle calculation}. 

At each snapshot, the filament center (in position and velocity) was defined as the local density peak. The resulting velocity projection for Filament 1 is shown in the upper panel of Figure \ref{fig:OFA 3D and flows}. Black dots represent the projected position of the sinks at each time step. Three distinct evolutionary phases were identified, delimited by the dashed pink and green horizontal lines, which are also indicated in the $ \cos(\theta_{\rm s,3D})$ heat map (top second panel). Before any preferred alignment emerges (below the pink dashed line), the inflow velocities toward the density peak are modest, on the order of $\sim\pm 2$ km s$^{-1}$. It can also be seen that the initially formed sinks move toward each other in a consistent manner, a behavior that reappears later in the evolution and typically occurs in regions of converging flow. During the stage in which a predominantly parallel alignment develops (between the pink and green dashed lines), the inflow velocity increase to $\sim\pm 6$ km s$^{-1}$, though this flow originates primarily from one side of the filament.

Finally, after $t=13.3$ Myr (above the green dashed line), when the alignment becomes predominantly perpendicular, accretion proceeds from both ends of the filament toward the densest region. This symmetric inflow may impart angular momentum to the central hub and promote the reorientation of the sink angular momenta. Furthermore, this reorientation can arise not only from the accretion of converging flows onto the sinks, but also from the contribution of orbital angular momentum during the formation of multiple systems by sink migration.

\begin{figure}
\centering \offinterlineskip
\includegraphics[width=\linewidth]{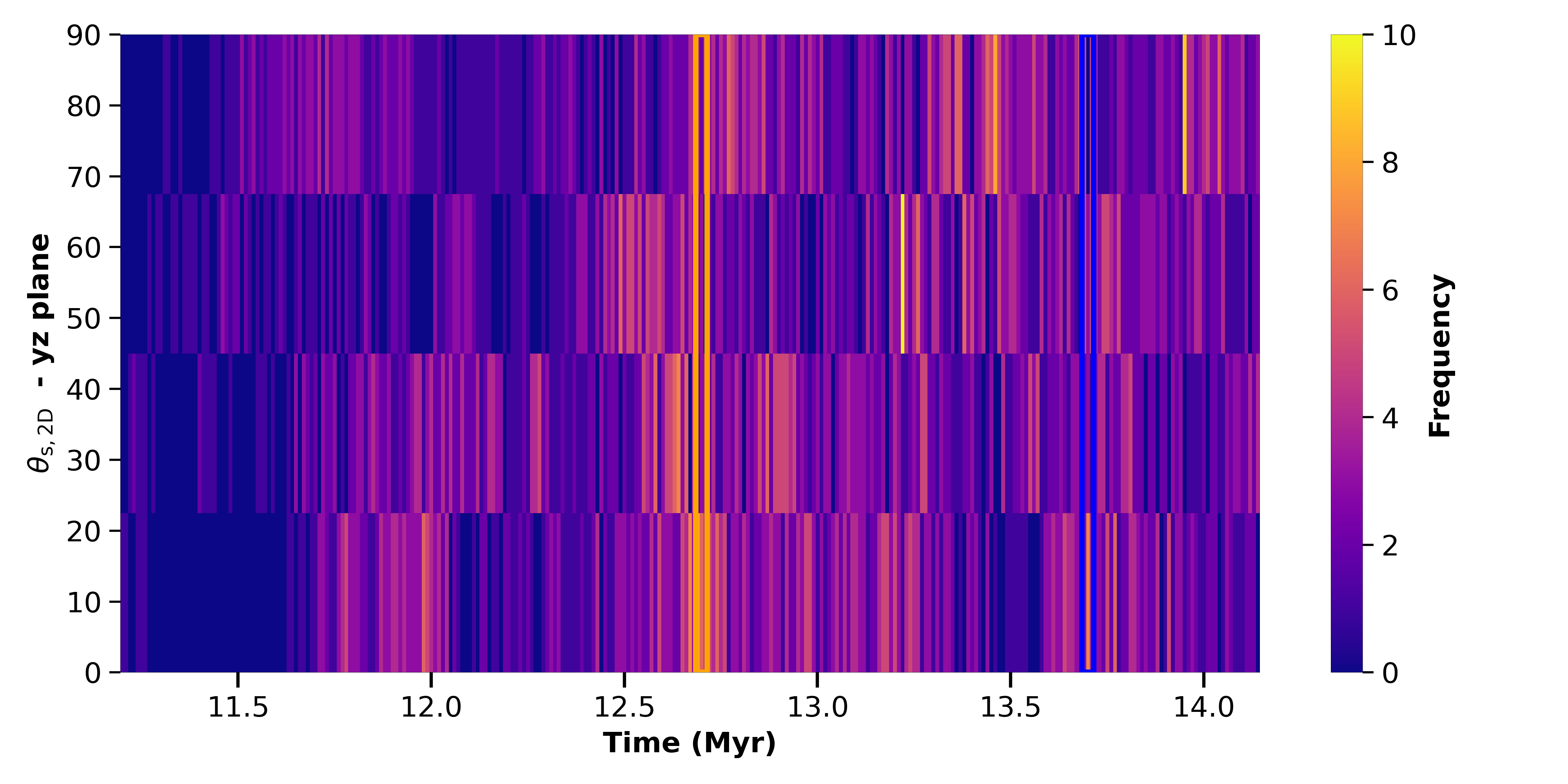}
\includegraphics[width=\linewidth]{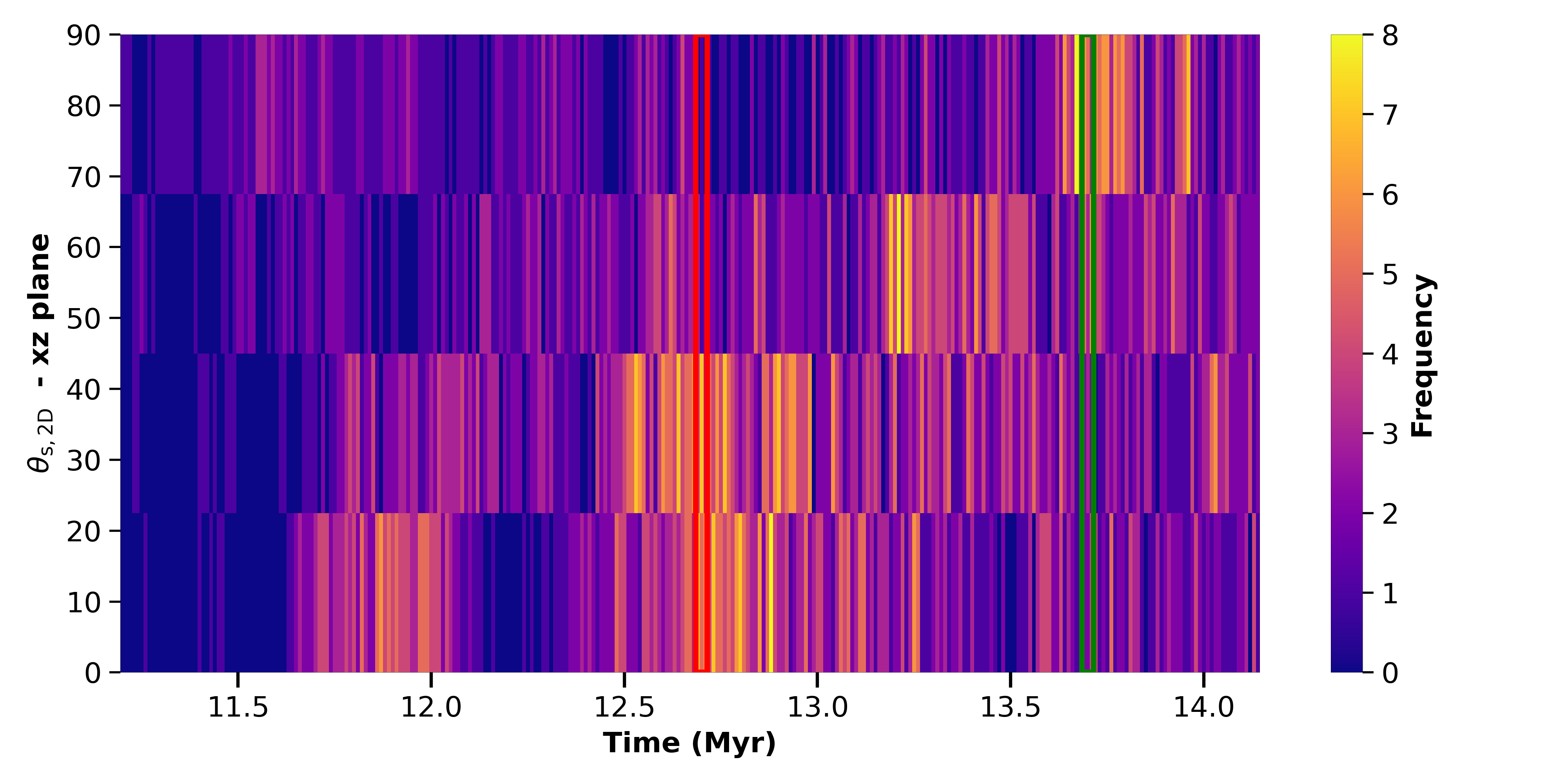}
\includegraphics[width=\linewidth]{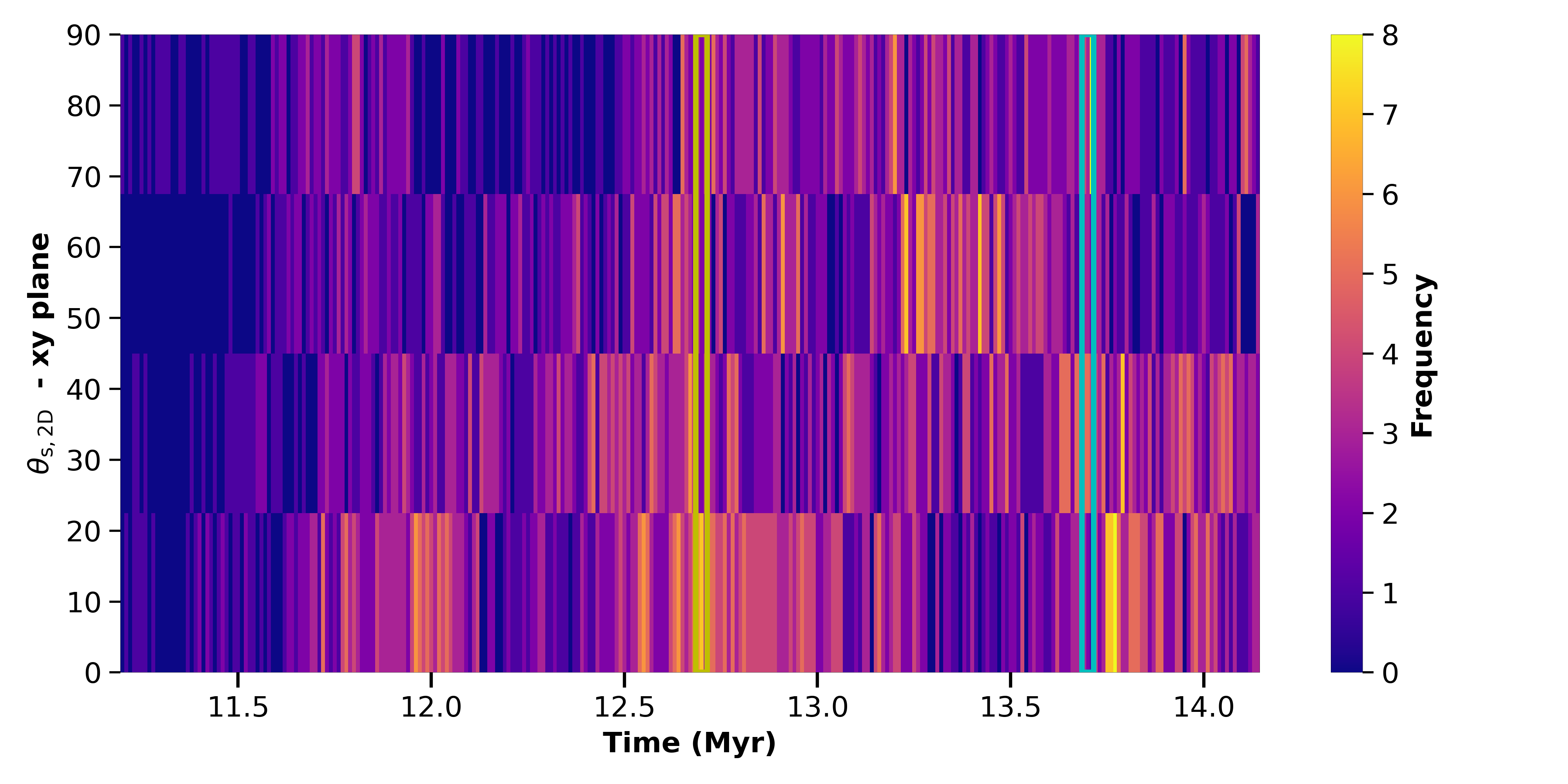}
 \caption{$\theta_{{\rm s,2D}}$ measured as the angle between the 2D projections of the angular momentum vectors of the sinks and the filament orientation onto the three coordinate planes, for the sample with no lifetime restriction in Filament 1. Cumulative distributions at times highlighted with warm ($t=12.7$ Myr) and cool ($t=13.7$ Myr) colors are shown individually in Figure \ref{fig:2D OFA examples fil1} following the same color code.}
 \label{fig:2D OFA evolution fil1}
\end{figure}

{In order to test whether this behavior is also recovered in the two-dimensional case, in Figure \ref{fig:2D OFA evolution fil1} we have projected the 3D distribution evolution of $\theta_{{\rm s,3D}}$ onto the three coordinate planes \footnote{Note that we present the results in terms of the angle itself rather than its cosine. In two dimensions, as is not the case in 3D, the distribution of angles between randomly oriented vector pairs is uniform, which provides the appropriate reference case against which we compare our measurements in order to identify any systematic deviation from randomness.}. We then take two snapshots at times $t=12.7$ (warm colors) and $t=13.7$ (cold colors) Myr which correspond to the times when the distribution appears mainly parallel and perpendicular respectively, and plot the cumulative distribution functions for $\theta_{{\rm s,2D}}$ at both times in Figure \ref{fig:2D OFA examples fil1}, following the same color pattern.

It can be seen that at time $t=12.7$ Myr the cumulative distributions tends to be above the solid black line, and therefore, closer to a parallel distribution (black dotted line). On the other hand, at time $t=13.7$ Myr, the cumulative distributions are positioned on and below the solid black line, with the projections on the $xz$ and $yz$ planes in green and blue respectively, being the ones that recover the distribution closest to the perpendicular distribution (black dashed line). Although the same behavior is reproduced in 3D and 2D particularly on the $xz$-plane, it is important to note, however, that the relatively small number of sinks leads to increased statistical noise in the cumulative histograms. Consequently, these results lack sufficient statistical data to draw strong conclusions regarding the 2D projection of the angle distribution. 

\begin{figure}
\centering \offinterlineskip
\includegraphics[width=\linewidth]{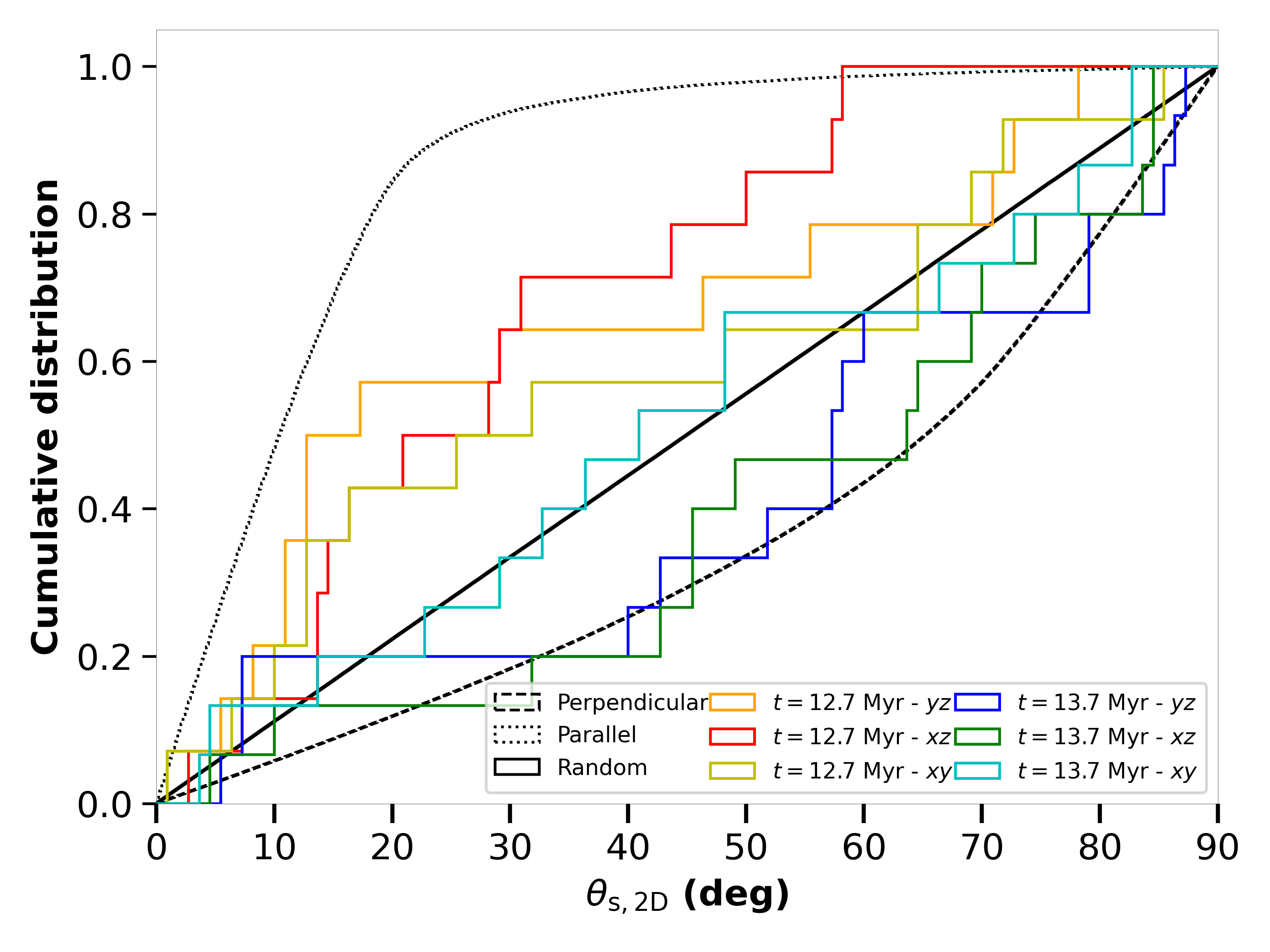}
 \caption{Cumulative histograms of the angle between the 2D projection of the angular momentum vector of the sinks and the filament orientation, $\theta_{{\rm s,2D}}$, at times $t=12.7$ (warm colors) and $13.7$ (cold colors) Myr (highlighted in Figure \ref{fig:2D OFA evolution fil1} for Filament 1) on the three coordinate planes $xy$, $xz$, and $yz$. No lifetime restrictions were applied.}
 \label{fig:2D OFA examples fil1}
\end{figure}

\subsubsection{Filament 2}
\label{subsubsec:Filament 2}

We also consider a second filament, identified as Filament 2, in which there does not appear to be a clear longitudinal flow, shown in the lower row of Figure \ref{fig:column density for individual filaments}. In this case, the sinks do not appear to be approaching each other and rather seem to remain in their place of origin. The two lower panels of Figure \ref{fig:OFA 3D and flows} presents the corresponding evolution of $cos(\theta_{{\rm s,3D}})$ distribution and the longitudinal velocity ($v_{||}$) for the filament axis obtained by fitting the spine points from \textsc{DisPerSE}, respectively, and using the same color scale for $v_{||}$ as in Filament 1. As expected, the angle distribution shows no significant preferential alignment throughout the $3$ Myr of evolution.

During the first $\sim 0.7$ Myr, a weak convergence towards the sinks is observed, whereas subsequently, the pattern does not appear to be evident, despite the increase in gas velocity along a preferential filament direction. Although these flows are substantially weaker than those in Filament 1, they may still contribute to the reorientation of the angular momentum vectors of some sinks, as suggested by the excess of angles near $90\degree$ after $t\sim10$ Myr in the bottom panel for the $cos(\theta_{{\rm s,3D}})$ distribution. However, the reduced number of sinks limits the statistical significance of this feature. 

\section{Discussion}
\label{sec:Discussion}

\subsection{How rapidly can gravity-driven longitudinal flows reorient sink angular momentum?}
\label{subsec:gravity action}

We have investigated the isolated role of gravity in driving longitudinal flows along filaments and its ability to produce a predominantly perpendicular alignment between sink angular momentum and the filament’s main axis, thereby testing the alignment mechanism proposed by \citet{Anathprindika.Withworth2008}. In our simulation, the angular momentum vectors of the sinks gradually reorient as shown in Figure \ref{fig:trayectory over OFA evolution}. Some sinks begin to oscillate near $90\degree$ by $t =\sim8$ Myr, whereas most of them tend to reorient by $t =\sim 10$ Myr. This timescales are broadly consistent with reported values for mass growth in filaments \citep[$\sim 1$-$2$ Myr,][]{Trevino-Morales+2019} and longitudinal collapse \citep[$\sim 1$-$4$ Myr,][]{Peretto+2014,Yuan+2020}. 

This raises the question of whether gravity alone can reorient sinks' angular momentum at a characteristic time within, for example, the lifetime of protostellar outflows. An additional consideration is that the sink tracking shown in Fig. \ref{fig:trayectory over OFA evolution} includes only sinks that survived long enough to follow their angular-momentum evolution. Sinks are removed from the sample once they merge with a more massive one. When sinks are tracked only during the first $0.5$ Myr of their lifetime, no preferential alignment is found between their angular momentum vectors and the filament axis. This suggests that gravity may not act rapidly enough to produce an observable reorientation within this early phase. However, the absence of a detectable signal may also reflect the global filament population. Only a minority of filaments exhibit strong longitudinal flows, while most resemble Filament 2 in the bottom panels of Fig. \ref{fig:column density for individual filaments}. If only a small subset of sinks undergo rapid reorientation, their contribution may be insufficient to dominate the overall angle distribution. Moreover, in filaments with significant longitudinal flows, such as Filament 1, sinks tend to migrate, increasing the likelihood that recently reoriented sinks merge and are therefore removed from the statistics. This would naturally reduce the number of angles near $90\degree$, further obscuring any early-time signature of preferential alignment.

A total of $399$ sinks associated with filaments were identified over the course of the simulation. By the final time considered, $142$ remained, implying that $64$\% were accreted by other sinks. Among the full sample, $43$ sinks formed with $\theta_{{\rm s,3D}}$ between $70\degree$ and $90\degree$ and remained within this range until their removal (by merging) or until the final snapshot. An additional $32$ sinks formed with $\theta_{{\rm s,3D}}$ outside this interval but later reoriented into the $70\degree-90\degree$ range and stayed there.

Top right panel of Figure \ref{fig:reorientation times} presents sink lifetimes, measured from formation until it leaves the statistics due to merging with a more massive sink or reaching the end of the simulation, vs reorientation times, defined as the time between sink formation and the moment when $\theta_{{\rm s,3D}}$ enters, and subsequently remains within, the $70\degree-90\degree$ interval for the $32$ sinks mentioned above. Color represents the simulation times at which these reorientations occur, and the dashed black line represents the identity line. These three quantities are presented individually in the histograms of the three remaining panels in the same figure. Most sinks reorient within $<0.5$ Myr, with a mode at $0.1$ Myr. The majority of reorientations take place at later times in the simulation, consistent with the increasing influence of gravity as some filaments develop stronger longitudinal flows.

This appears to suggest that, once longitudinal flows are established, they may be capable of reorienting the angular momentum of the sink within $0.5$ Myr. However, such flows occur only in a subset of filaments, and at the adopted resolution they also promote sink migration and subsequent merging, as can be seen from their short lifespan. In fact, the clustering of sinks near the identity line (dashed black line) suggests that once they undergo significant reorientation, they are rapidly accreted. As a consequence, the fraction of sinks that undergo early and sustained reorientation is not large enough to generate a clear statistical signal in the full sample.

\begin{figure*}
\centering \offinterlineskip
\includegraphics[width=0.35\linewidth]{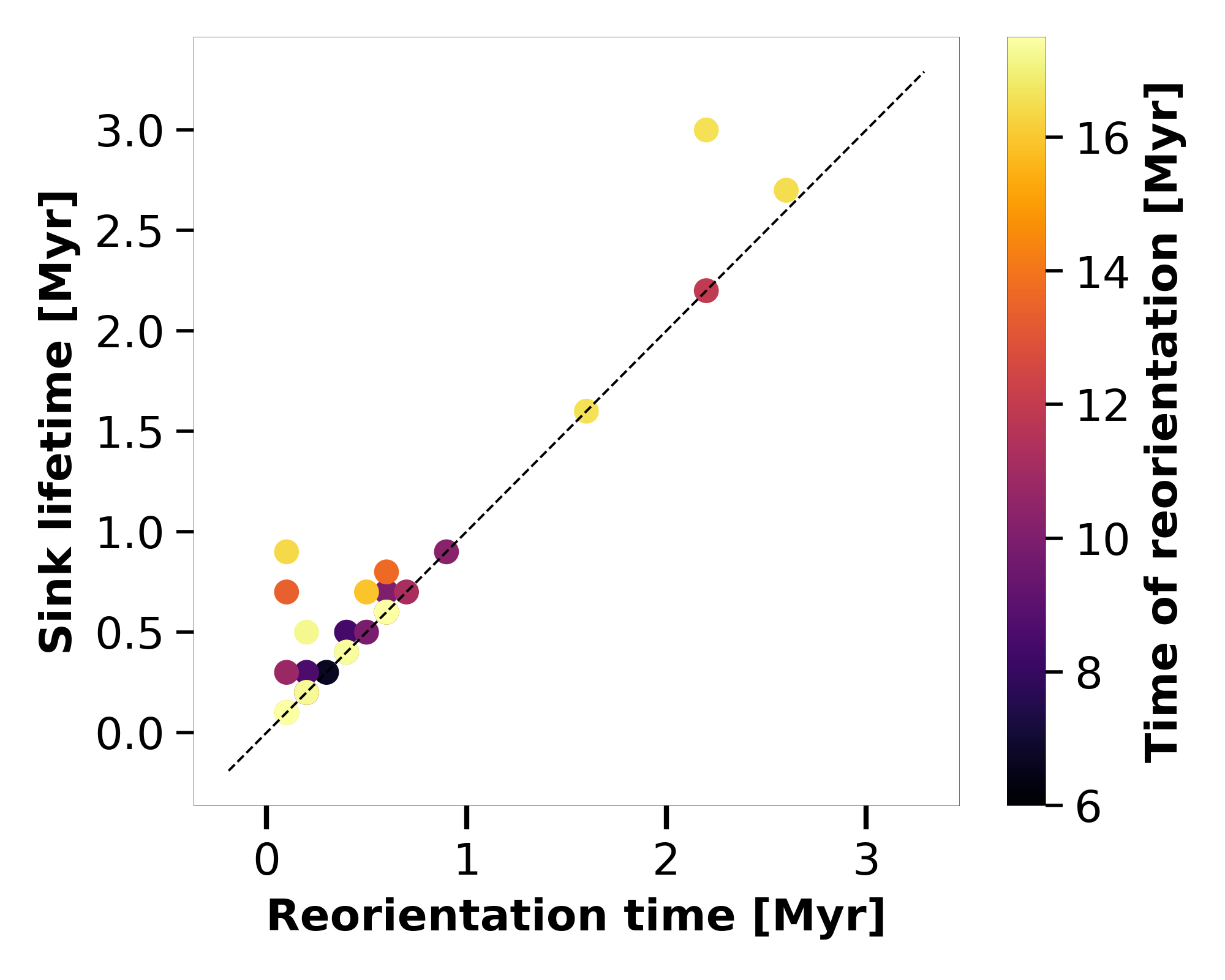}
\includegraphics[width=0.35\linewidth]{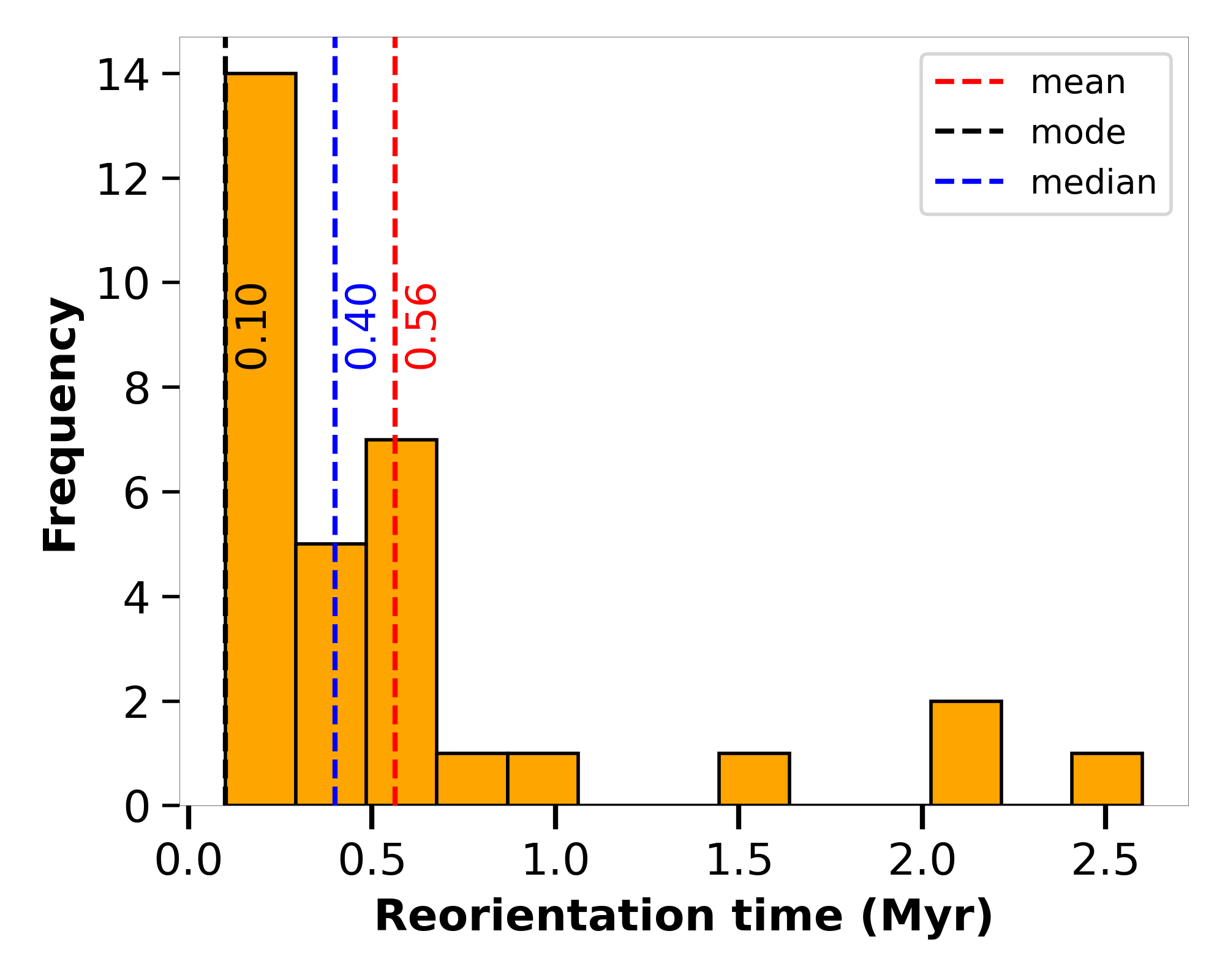}
\includegraphics[width=0.35\linewidth]{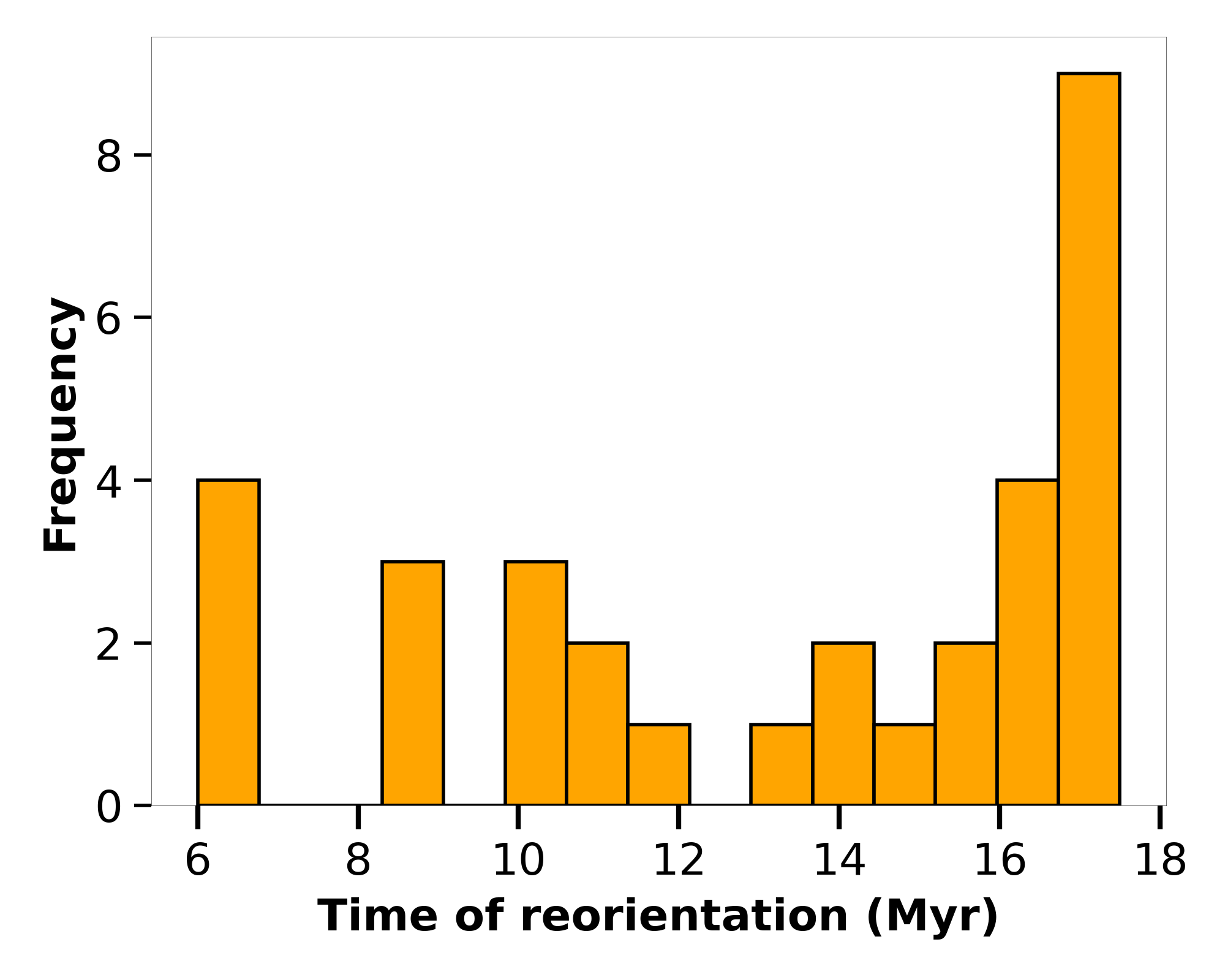}
\includegraphics[width=0.35\linewidth]{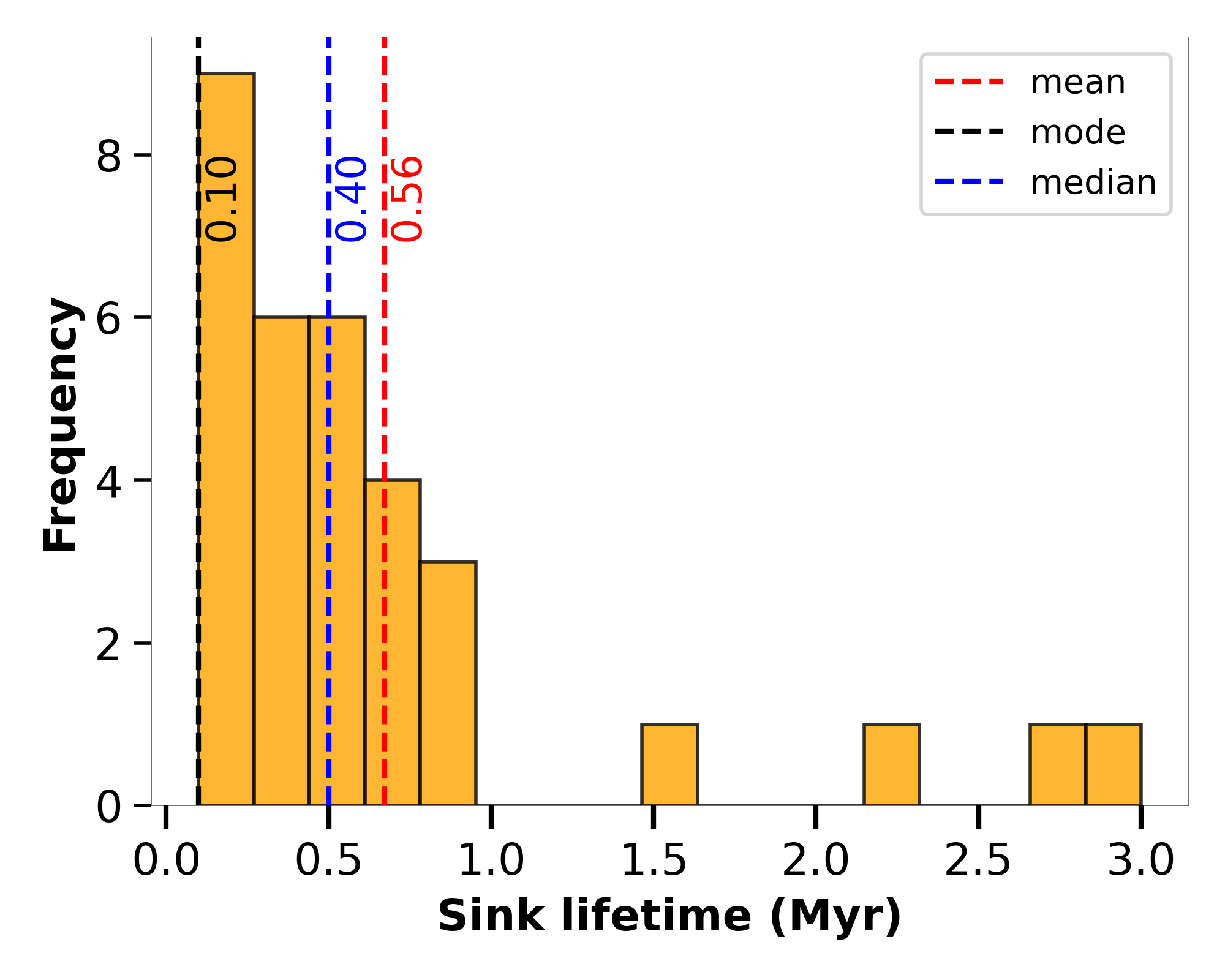}
 \caption{{\it Top left:} Lifetimes of a subsample of 32 sinks (see text), measured from formation until it leaves the statistics due to merging with a more massive sink or reaching the end of the simulation, vs reorientation times, defined as the time between sink formation and the moment when $\theta_{{\rm s,3D}}$ enters, and subsequently remains within, the $70\degree-90\degree$ interval. Color represents the simulation times at which these reorientations occur. The dashed black line represents the identity line. These three quantities are presented individually in the histograms of the three remaining panels.}
 \label{fig:reorientation times}
\end{figure*}


\subsection{3D perpendicular angle fraction for 2D perpendicular alignment detection}
\label{subsec:statistical model}

Throughout this work, we have shown that, while the distribution of $\theta_{\rm s,3D}$ is predominantly perpendicular at later times in the simulation for the full sink sample, its 2D projection appears closer to random. As discussed above, this arises from the fact that the angle distribution between two randomly oriented 3D vectors is intrinsically non-uniform and peaks near $90^{\degree}$, whereas its 2D projection yields a uniform distribution. Consequently, detecting a perpendicular alignment in the cumulative histograms of the 2D projection requires an excess of angles near $90^{\degree}$ relative to the random expectation. To quantify this requirement, we carry out a Monte-Carlo forward modeling analysis to determine the minimum fraction $f$ of 3D angles in the $70^{\degree}-90^{\degree}$ range needed for the projected 2D cumulative distribution function (CDF) to be statistically distinguishable from a random distribution, thereby enabling the identification of a perpendicular alignment.

We define a sample of $\theta_{\rm s,2D}$ measurements composed of a mixture of two subsamples: one corresponding to a 3D random distribution (hereafter, {\it null} sample), and another corresponding to a perpendicular distribution, generated from 3D angles in the range $70^{\degree}-90^{\degree}$. In this way, we can represent the CDF of the mixture ($F_{\rm mix}$) as a linear combination such that
\begin{equation}
    F_{\rm mix} (\theta;f)=(1-f)F_{\rm null} (\theta) + f F_{\rm perp} (\theta).
    \label{eq:F_mix}
\end{equation}
To compare two CDFs, $G$ and $H$, we employ the Kolmogorov-Smirnov (KS) statistic, defined as the maximum vertical separation between them. Formally,
\begin{equation}
    S(G,H) =  \sup_{\theta} | G(\theta) - H(\theta)| .
    \label{eq:Delta F}
\end{equation}
We aim to compare the full (mixed) sample with the null (random) sample in order to determine the value of $f$ that maximizes their difference. Identifying these two samples as $F_{\rm mix} \rightarrow G$ and $F_{\rm perp} \rightarrow H$ in equation \eqref{eq:Delta F}, and using equation \eqref{eq:F_mix}{}, we have
\begin{align}
     G(\theta) - H(\theta) &= F_{\rm mix} (\theta;f) - F_{\rm null} (\theta)   \notag\\
    &=  (1-f)F_{\rm null} (\theta)  \notag\\
    & + f F_{\rm perp} (\theta) - F_{\rm null} (\theta) \notag \\
    & =  f[ F_{\rm perp} (\theta) - F_{\rm null} (\theta)].
\end{align}
Then, 
\begin{align}
     S(f) &=f \cdot  \sup_{\theta} |  F_{\rm perp} (\theta) - F_{\rm null} (\theta)|    \notag\\
    &= f D_{\rm max},
    \label{eq:fD}
\end{align}
with $D_{\rm max}$ being the maximum difference between the 2D CDFs of the perpendicular and null samples.

In practice, we work with empirical CDFs (ECDFs), constructed from a finite sample of N measurements or events, which approximate the true underlying CDF. ECDFs exhibit sampling fluctuations that typically scale as $N^{-1/2}$ \citep{Vaart1998}. Therefore, to detect a significant difference between two CDFs, the intrinsic noise in the ECDFs must exceed:
\begin{equation}
    S(f) \geq \frac{C}{\sqrt{N_{\theta}}}. 
    \label{eq: D noise}
\end{equation}
In this equation, $C$ is a constant that depends on the decision method. For the case of a single-sample KS test at level $\alpha=0.05$, the classical constant \citep{Smirnov1948} is
\begin{equation}
C = K_{\alpha} = 1.36.
\end{equation}
Then, from equations \eqref{eq: D noise} and \eqref{eq:fD},
\begin{equation}
    f(N) = \frac{K_{\alpha}}{D_{\rm max} \sqrt{N_{\theta
    }}}.
    \label{eq:F(N)}
\end{equation}
$D_{\rm max}$ does not have a known analytical form, but can be estimated empirically through Monte-Carlo experiments. To do so, we compute the angle between pairs of randomly oriented 3D vectors and define two samples: a null sample considering all angles, and a perpendicular sample restricted to angles between $70^{\degree}$ and $90^{\degree}$. The generation of the 3D random vectors and their 2D projections was computed following \citet{Stephens+2017}. As expected, the larger the angle samples, the better the value of $D_{\rm max}$ will be constrained. We then evaluate the maximum vertical separation between their CDFs in 2D obtaining
\begin{equation}
    D_{\rm max} = 0.23233.
\end{equation}
Then, from equation \eqref{eq:F(N)},
\begin{equation}
    f(N) = \frac{K_{\alpha}}{D_{\rm max} \sqrt{N_{\theta}}} = \frac{1.36}{0.23233 \sqrt{N_{\theta}}} = \frac{5.854}{\sqrt{N_{\theta}}}.
    \label{eq: f(N) final}
\end{equation}
It is worth mentioning that this $f$ represents a lower limit on the required fraction of angles close to $90\degree$, since the null sample already has an intrinsic majority of these angles. Therefore, it should be understood as the excess of angles above the null distribution.

We can then apply this framework to observational data. For instance, \citet{Kong+2019} report OFA measurements for approximately $60$ outflows, finding a distribution clearly dominated by perpendicular alignments. Setting $N_{\theta} = 60$ in equation \eqref{eq: f(N) final}, we find that an excess of at least $76\%$ of 3D angles within $70^{\degree}$ and $90^{\degree}$ is required for the projected CDF to be statistically distinguishable from a random distribution. In the simulation analyzed here, at time $t=17.5$ Myr, there is a population of $142$ filament-associated sinks, of which $42.2\%$ fall within the $70^{\degree}-90^{\degree}$ range. However, substituting $N_{\theta}=142$ into equation  \eqref{eq: f(N) final}, we find that we would need an excess of at least $49\%$ of angles in this range for the distribution of projected 2D angles to be statistically distinguishable from the random distribution. We suggest that this could explain why the cumulative histogram $\theta_{\rm s,2D}$ appears indistinguishable from a random distribution. Due to intrinsic noise, we cannot give a statistically significant conclusion for angle samples with $N_{\theta}<35$.  In Figure \ref{fig:f(N)}, we show the dependence of $f(N_{\theta})$ on $N_{\theta}$. The dashed red vertical line represents the minimum number of measurements for which a statistically significant conclusion can be drawn. The value of $N_{\theta}=60$ is highlighted in cyan for comparison with \citet{Kong+2019}.

\begin{figure}
\centering \offinterlineskip
\includegraphics[width=\linewidth]{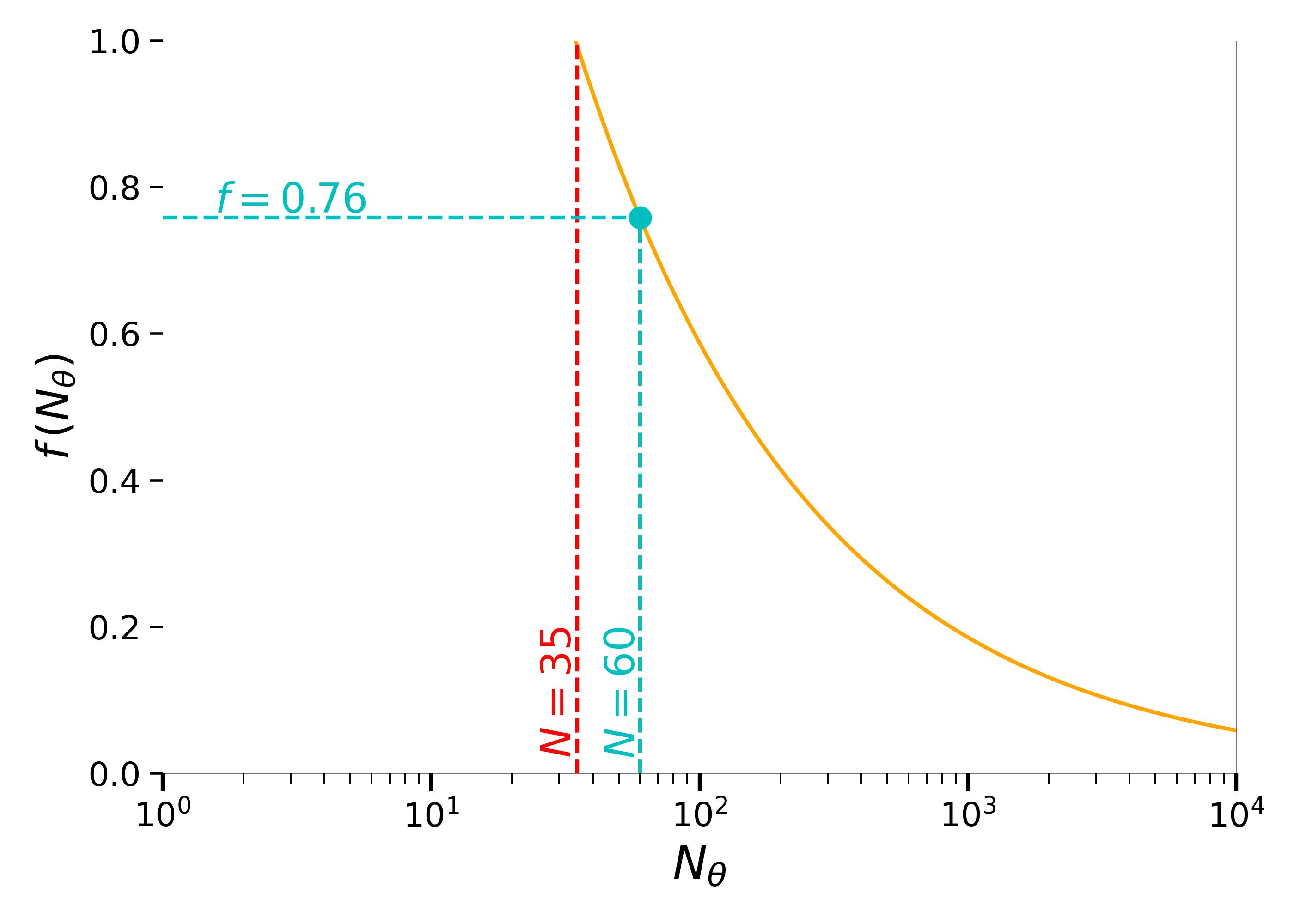}
\caption{Minimum 3D perpendicular angle fraction for 2D perpendicular alignment detection, $f (N_{\theta})$, as a function of the number of angles, $N_{\theta}$, according to equation \eqref{eq: f(N) final}. $N_{\theta}=35$ in red represents the minimum number of measurements for which a statistically significant estimate above the intrinsic noise can be obtained. The corresponding value for $N_{\theta} = 60$ are shown in cyan for comparison with \citet{Kong+2019}.}
 \label{fig:f(N)}
\end{figure}






\section{Caveats}
\label{sec:Caveats}
Three main limitations of the present study should be acknowledged, which also point toward promising directions for future work. First, the numerical resolution of the full-box simulation remains far above the prestellar-core, first-core, disk, and outflow-launching scales. Thus, the sink particles analyzed here should not be interpreted as individual protostars or as direct proxies for protostellar outflow directions. Instead, they trace unresolved collapsing fragments on low-density core/clump scales. The angular momentum measured in this work therefore represents the angular momentum delivered to unresolved fragments by filament-scale gas motions. Additional torques during the subsequent collapse, including gravitational, magnetic, and disk-scale torques, may alter both its magnitude and direction. 

Nevertheless, the two zoom-in simulations presented in this work provide a useful first convergence check. In both cases, the large-scale filament morphology and the qualitative behavior of the $\theta_{\rm s,3D}$ distribution are preserved at higher resolution, while the refined runs mainly reveal additional internal structure on smaller scales. This raises the question of whether small-scale turbulent fluctuations in the diffuse medium, where the spatial resolution is naturally lower, as in any numerical method with adaptive resolution, contribute significantly to the net rotation of clumps within the filament, and ultimately affect the angular momentum orientation of the sinks formed within them. For a standard turbulent spectrum, such as Kolmogorov or Burgers turbulence, the characteristic velocity difference across a scale $L$ follows $\delta v \sim L^\alpha$, with $\alpha = 1/3$ for Kolmogorov, and $\alpha = 1/2$ for Burgers turbulence. Thus, larger spatial scales are associated with larger velocity differences and more spatially coherent motions. We therefore expect a coherent net rotation of the clump to be driven mainly by gas motions on scales comparable to, or larger than, the clump itself, rather than by smaller-scale turbulent fluctuations.

Second, the simulation does not include magnetic fields. This is an important limitation, since magnetic fields can modify the angular momentum budget through magnetic braking and non-ideal MHD effects, and may affect the relative orientation between the filament, the angular momentum vector, and the outflow-launching direction \citep{Cortes+2006,Matsumoto+2006,Beuther+2010,Hull+2014,Lee+2017,Galametz+2018,Galametz+2020,Commercon+2022,Encalada+2024,Huang+2024}. The present work intentionally isolates the hydrodynamical and gravitational contribution associated with filament-scale longitudinal flows. A complete comparison with outflow orientations requires higher-resolution magnetized simulations capable of following the collapse from filament scales to dense-core and disk scales.

Finally, the inclusion of stellar feedback is expected to hinder the formation and long-term maintenance of longitudinal accretion flows, potentially limiting their ability to redistribute angular momentum efficiently. Under such conditions, it becomes less straightforward to assess whether the angular momentum reorientation timescales of the sinks are sufficiently short to operate before feedback disrupts the accretion flows. As we incorporate feedback processes into our simulations, we aim to examine this interplay in greater detail in future work.


\section{Conclusions}
\label{sec:Conclusions}

We analyzed filaments in a simulation of molecular cloud formation in a warm, neutral medium, including decaying turbulence, gravity, cooling and heating, and sink formation. The 3D and projected 2D angles between sink angular momentum and filament orientation were measured  over $\sim 11$ Myr, considering the full sink sample and those sinks formed in two individual filaments with distinct dynamics. We further examined the relation between predominantly perpendicular alignment and the development of longitudinal flows, the reorientation times of sink subsamples, and the fraction of 3D perpendicular angles required to distinguish a perpendicular distribution from a random one in 2D cumulative histograms.

Our conclusions can be summarized as follows:

\begin{itemize}
    \item During the first $\sim 5$ Myr after the formation of the first sink, the 3D distributions of angles between sink angular momentum and the local filament direction are consistent with randomness, showing no preferential orientation. At later times, a clear excess of angles closer to $90\degree$ emerges in both the angle and its cosine, indicating a transition from an initially random configuration to a predominantly perpendicular alignment.

    \item The evolution of the 2D angle distribution, projected onto the three coordinate planes, remained consistent with randomness throughout the analyzed period. Although an excess of 3D angles in the $70\degree-90\degree$ range is apparent in the cosine distribution, it is insufficient to distinguish the distribution from random in 2D cumulative histograms. 
    
    \item Additionally, using a simple statistical toy model, we estimate that the minimum detectable fraction of intrinsically perpendicular alignment scales as $f(N_\theta)=5.854/\sqrt{N_\theta}$. Under our assumptions, samples with fewer than $\sim35$ angles are insufficient to distinguish even a fully perpendicular 3D population from a random one after projection. This highlights how strongly projection and limited sample size can erase intrinsic alignment signatures in 2D.
    
    \item Analysis of the orientation of SPH particle velocity vectors relative to the filament spine shows that at late times, when gravity likely dominates the dynamics of dense regions, an excess of angles near zero emerges, indicating velocities aligned with the filament axis. This suggests the development of longitudinal flows, coinciding with the appearance of a predominantly perpendicular distribution of angles between sink angular momentum and the filament axis.

    \item The study of two individual filaments indicates that convergent flows toward the density peak are associated with a predominantly perpendicular distribution of angles between sink angular momentum and the filament axis, whereas weak or absent flows correspond to a more random distribution. Additionally, the occurrence of a predominantly parallel angle distribution in regions with unidirectional (non-convergent) flows warrants further investigation to clarify any potential connection.

    \item Individual sink tracking indicates that the emergence of a predominantly perpendicular angle distribution between sinks' angular momentum vector and filament's direction arises from reorientation during the course of the simulation. However, due to the simulation’s sink merging criterion, sinks forming with $\theta_{{\rm s,3D}}$ near $90\degree$ in regions already dominated by gravity may be undercounted, as they merge rapidly. Indeed, sinks that reorient to $70\degree-90\degree$ predominantly do so later in the simulation, coinciding with the development of longitudinal flows, yet they also have short lifetimes before merging. 
    
    \item  Reorientations of sinks' angular momentum vector can occur as quickly as $0.1$ Myr, suggesting that once gravity dominates filament dynamics, this effect can have a significant impact on setting the initial conditions for the subsequent formation and evolution of outflows.

\end{itemize}


\appendix

\section{Angular momentum treatment during accretion and merging of sinks}
\label{Appe:AM tratment}

In \textsc{Phantom}, sink particles carry a spin angular momentum ${\bf S}_{\rm s}$, stored independently from their orbital angular momentum. When gas particles are accreted, or when two sinks merge, the code updates the sink mass, position, velocity, and spin so that the total angular momentum is conserved. 

For the accretion of gas particles onto a sink, the updated sink mass, position, and velocity are
\begin{equation}
 M' = M_{\rm s} + \sum_i m_i ,
\end{equation}
\begin{equation}
 {\bf r}' =
 \frac{M_{\rm s}{\bf r}_{\rm s}+\sum_i m_i{\bf r}_i}{M'},
\end{equation}
\begin{equation}
 {\bf v}' =
 \frac{M_{\rm s}{\bf v}_{\rm s}+\sum_i m_i{\bf v}_i}{M'} ,
\end{equation}
where \(M_{\rm s}\), \({\bf r}_{\rm s}\), and \({\bf v}_{\rm s}\) are the
pre-accretion sink mass, position, and velocity, and \(m_i\), \({\bf r}_i\), and \({\bf v}_i\) are the masses, positions, and velocities of the accreted gas particles. The spin of the updated sink is then
\begin{equation}
 {\bf S}' =
 {\bf S}_{\rm s}
 + M_{\rm s}{\bf r}_{\rm s}\times{\bf v}_{\rm s}
 + \sum_i m_i{\bf r}_i\times{\bf v}_i
 - M'{\bf r}'\times{\bf v}' .
\end{equation}
Here, the first term on the right-hand side is the pre-existing internal spin of the sink. The next two terms are the orbital angular momenta, measured about the origin, of the original sink and of the accreted gas particles, respectively. The final term subtracts the orbital angular momentum of the updated sink center of mass, leaving only the unresolved internal angular momentum to be stored as the spin of the new sink. This subtraction is necessary because the first three terms include the total angular momentum measured about the origin; removing the orbital contribution of the new center of mass ensures that the total angular momentum, ${\bf J} = M'{\bf r}'\times{\bf v}' + {\bf S}'$, is conserved.

For the merger of two sinks, labeled 1 and 2, the new sink has
\begin{equation}
 M' = M_1+M_2 ,
\end{equation}
\begin{equation}
 {\bf r}' =
 \frac{M_1{\bf r}_1+M_2{\bf r}_2}{M'},
\end{equation}
\begin{equation}
 {\bf v}' =
 \frac{M_1{\bf v}_1+M_2{\bf v}_2}{M'} .
\end{equation}
The resulting spin is
\begin{equation}
 {\bf S}' =
 {\bf S}_1+{\bf S}_2
 + M_1{\bf r}_1\times{\bf v}_1
 + M_2{\bf r}_2\times{\bf v}_2
 - M'{\bf r}'\times{\bf v}' .
\end{equation}
This can also be written in center-of-mass form as
\begin{equation}
 {\bf S}' =
 {\bf S}_1+{\bf S}_2
 + M_1({\bf r}_1-{\bf r}')\times({\bf v}_1-{\bf v}')
 + M_2({\bf r}_2-{\bf r}')\times({\bf v}_2-{\bf v}') .
\end{equation}
Thus, during a sink merger, the orbital angular momentum of the unresolved
binary is converted into spin angular momentum of the merged sink. This
bookkeeping ensures conservation of the total angular momentum,
\begin{equation}
 {\bf J}_{\rm tot}
 =
 \sum_k m_k{\bf r}_k\times{\bf v}_k
 + \sum_{\rm sinks}{\bf S}_{\rm sink},
\end{equation}
up to numerical accuracy. The sink-particle prescription follows the general approach of \citet{Bate+1995}, while the conservative angular-momentum bookkeeping described above corresponds to the implementation in \textsc{Phantom}.

\section{Examples of sink angular momentum evolution}
\label{Appe:Sink trajectories}

Top panel of Figure \ref{fig:trayectory over OFA evolution} shows the evolution of the cosine of the angles between the angular momentum vector (spin) and the host filament for three sinks selected as those that remain active for most of the simulation. In the lower panel, the trajectories were smoothed using a Gaussian time kernel, in which nearby points are weighted more strongly and local outliers are iteratively down-weighted. As can be seen, during the first few million years of tracking, there are significant fluctuations in the measurements of $cos(\theta_{{\rm s,3D}})$, possibly due to the early formation of filaments, or accretion processes onto the sinks. From $\sim t=10$ Myr onwards, a trend emerges towards oscillations near $90\degree$ ($cos(\theta_{{\rm s,3D}}) \sim 0$), and therefore explaining the high density of angles in this region. One possible explanation for this reorientation could be the development of longitudinal flows as gravity begins to exert a stronger influence on the filaments dynamics, and therefore, favoring the scenario proposed by \citet{Anathprindika.Withworth2008}. 
\begin{figure*}
\centering \offinterlineskip
\includegraphics[width=\linewidth]{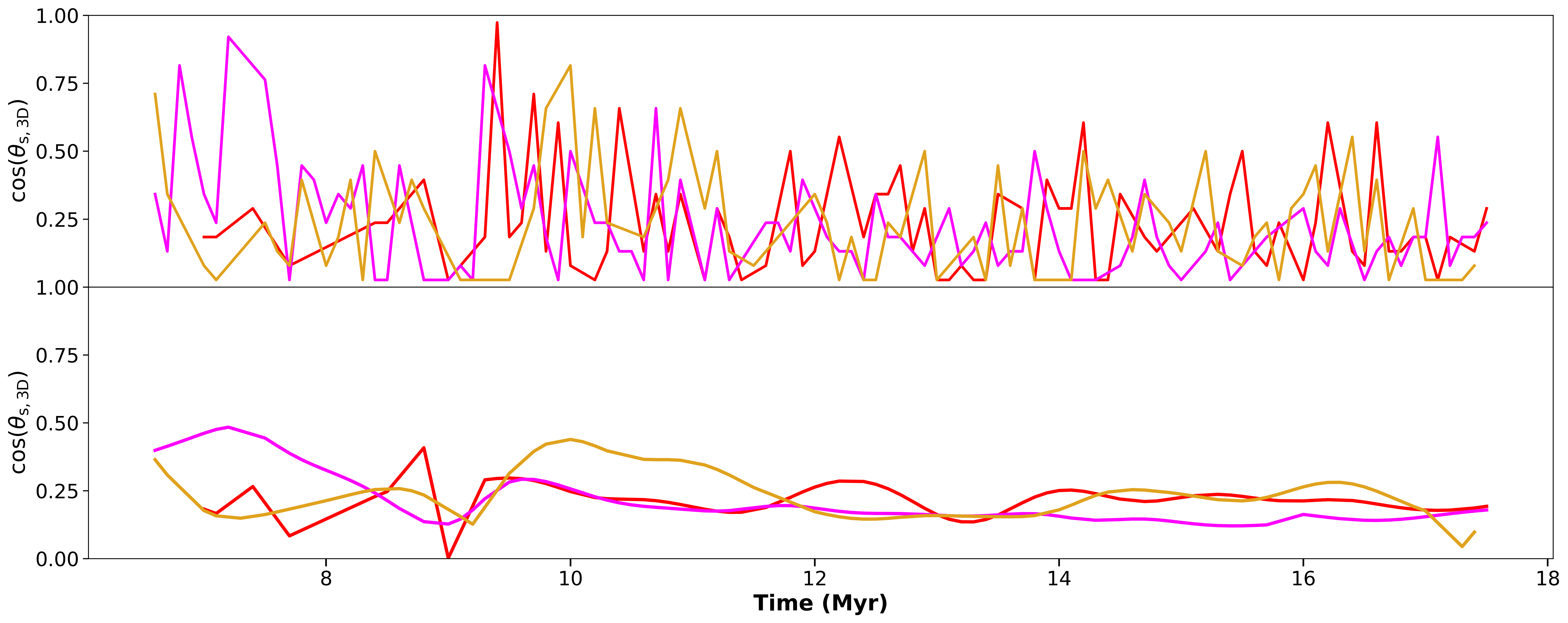}
 \caption{Top panel shows the evolution of the cosine of the angles between the angular momentum vector (spin) and the host filament for three sinks (differentiated by color) from their time of formation. In the bottom panel the trajectories were smoothed using a Gaussian kernel. At early times, a drastic oscillation is observed, while at later times the oscillation is around angles closer to $90\degree$ ($cos(\theta_{{\rm s,3D}}) \sim 0$).}
 \label{fig:trayectory over OFA evolution}
\end{figure*}

\section{Filament 1 comparison at low and high resolution}
\label{Appe:Fil 1 compare resolution}

As a convergence test, in the leftmost panel of Fig. \ref{fig:Fil 1 compare resolution} we show the splitting time at $t = 10$ Myr for Filament 1 (analyzed in Section \ref{subsubsec:Filament 1}), followed by a comparison between low and high resolution at $t = 12.5$ Myr. The region where the splitting took place encompasses an area larger than the one shown. The large-scale morphology is preserved at high resolution, while more small-scale structure develops, as illustrated in the right panel, which shows a zoom-in of the cyan box on the high-resolution panel. Specifically, a section of the filament is shown exhibiting misaligned longitudinal flows directed toward the sink at $x \sim 111$ pc.

\begin{figure*}
\centering \offinterlineskip
\includegraphics[width=\linewidth]{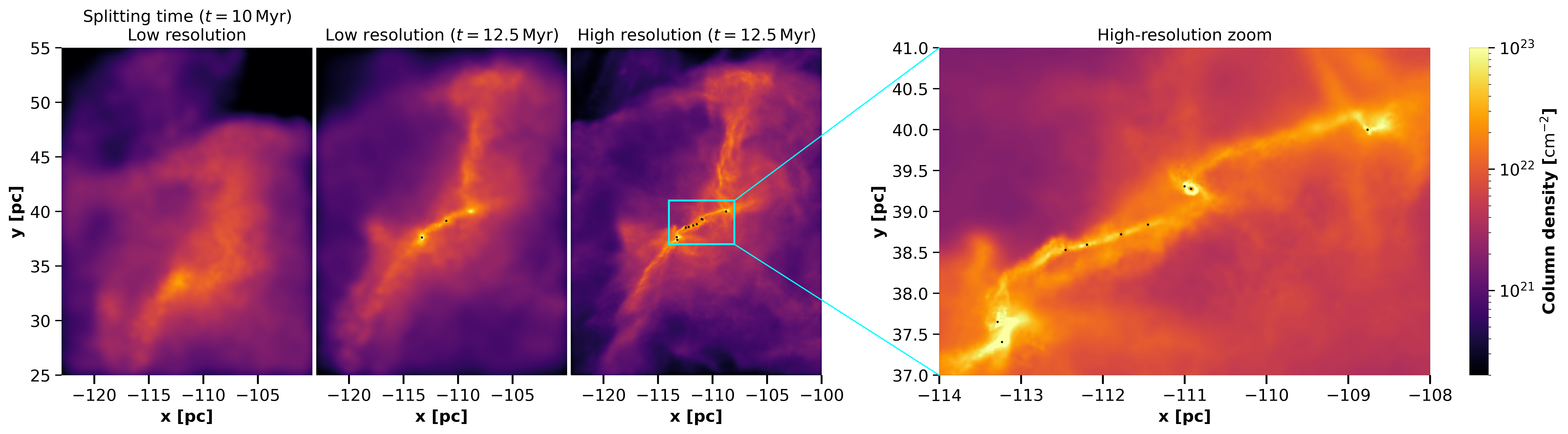}
 \caption{Column-density maps of Filament 1 at the splitting time ($t=10$ Myr), and the comparison at time $t=12.5$ at low and high resolution. The right panel shows a zoomed-in view of the region outlined in cyan in the high-resolution image. Black dots indicate active sink particles.
}
 \label{fig:Fil 1 compare resolution}
\end{figure*}


\begin{acknowledgments}
We thank the anonymous referee for helpful comments.
\end{acknowledgments}

\software{Phantom, \citep{Price+2018}, Numpy \citep{numpy}, \citep{Robitaille12}, Matplotlib \citep{matplotlib}, Sarracen \citep{Harris.Tricco2023}, Plonk \citep{Mentiplay2019}, Scipy \citep{SciPy-NMeth}, scikit-learn \citep{scikit-learn}}


\bibliography{biblio,ref}
\bibliographystyle{aasjournalv7}

\end{document}